\documentclass[sigconf]{acmart}

\usepackage{multirow}
\usepackage{bm}
\usepackage[ruled,linesnumbered]{algorithm2e}
\usepackage{amsmath}

\AtBeginDocument{
	\providecommand\BibTeX{{
			\normalfont B\kern-0.5em{\scshape i\kern-0.25em b}\kern-0.8em\TeX}}}

\copyrightyear{2024} 
\acmYear{2024} 
\setcopyright{acmlicensed}
\acmConference[MM '24]{Proceedings of the 32nd ACM International Conference on Multimedia}{October 28-November 1, 2024}{Melbourne, VIC, Australia}
\acmBooktitle{Proceedings of the 32nd ACM International Conference on Multimedia (MM '24), October 28-November 1, 2024, Melbourne, VIC, Australia}
\acmDOI{10.1145/3664647.3681512}
\acmISBN{979-8-4007-0686-8/24/10}

\begin{document}
	\title{Perceptual-Distortion Balanced Image Super-Resolution is a Multi-Objective Optimization Problem}
	
	\author{Qiwen Zhu}
	\orcid{0009-0008-5984-4941}
	\affiliation{
		\institution{State Key Lab of MIIPT, Huazhong University of Science and Technology}
		\city{Wuhan}
		\country{China}
	}
	\email{zhuqiwen@hust.edu.cn}
	\author{Yanjie Wang}
	\orcid{0000-0001-7352-6183}
	\affiliation{
		\institution{School of AIA, Huazhong University of Science and Technology}
		\city{Wuhan}
		\country{China}
	}
	\email{aiawyj@hust.edu.cn}
	\author{Shilv Cai}
	\orcid{0000-0002-4037-4555}
	\affiliation{
		\institution{School of AIA, Huazhong University of Science and Technology}
		\city{Wuhan}
		\country{China}
	}
	\email{caishilv@hust.edu.cn}
	\author{Liqun Chen}
	\orcid{0000-0001-6345-5713}
	\affiliation{
		\institution{School of AIA, Huazhong University of Science and Technology}
		\city{Wuhan}
		\country{China}
	}
	\email{chenliqun@hust.edu.cn}
	\author{Jiahuan Zhou}
	\orcid{0000-0002-3301-747X}
	\affiliation{
		\institution{Wangxuan Institute of Computer Technology, Peking University}
		\city{Beijing}
		\country{China}
	}
	\email{jiahuanzhou@pku.edu.cn}
	\author{Luxin Yan}
	\orcid{0000-0002-5445-2702}
	\affiliation{
		\institution{State Key Lab of MIIPT, Huazhong University of Science and Technology}
		\city{Wuhan}
		\country{China}
	}
	\email{yanluxin@hust.edu.cn}
	\author{Sheng Zhong}
	\orcid{0000-0003-2865-8202}
	\affiliation{
		\institution{State Key Lab of MIIPT, Huazhong University of Science and Technology}
		\city{Wuhan}
		\country{China}
	}
	\email{zhongsheng@hust.edu.cn}
	\author{Xu Zou}
	\orcid{0000-0002-0251-7404}
	\authornote{Corresponding author.}
	\affiliation{
		\institution{State Key Lab of MIIPT, Huazhong University of Science and Technology}
		\city{Wuhan}
		\country{China}
	}
	\email{zoux@hust.edu.cn}

	\begin{abstract}
		Training Single-Image Super-Resolution (SISR) models using pixel-based regression losses can achieve high distortion metrics scores (\emph{e.g.}, PSNR and SSIM), but often results in blurry images due to insufficient recovery of high-frequency details. Conversely, using GAN or perceptual losses can produce sharp images with high perceptual metric scores (\emph{e.g.}, LPIPS), but may introduce artifacts and incorrect textures. Balancing these two types of losses can help achieve a trade-off between distortion and perception, but the challenge lies in tuning the loss function weights. To address this issue, we propose a novel method that incorporates Multi-Objective Optimization (MOO) into the training process of SISR models to balance perceptual quality and distortion. We conceptualize the relationship between loss weights and image quality assessment (IQA) metrics as black-box objective functions to be optimized within our Multi-Objective Bayesian Optimization Super-Resolution (MOBOSR) framework. This approach automates the hyperparameter tuning process, reduces overall computational cost, and enables the use of numerous loss functions simultaneously. Extensive experiments demonstrate that MOBOSR outperforms state-of-the-art methods in terms of both perceptual quality and distortion, significantly advancing the perception-distortion Pareto frontier. Our work points towards a new direction for future research on balancing perceptual quality and fidelity in nearly all image restoration tasks. The source code and pretrained models are available at:  \href{https://github.com/ZhuKeven/MOBOSR}{https://github.com/ZhuKeven/MOBOSR}.
	\end{abstract}
	
	
	
	
	\begin{teaserfigure}
		\centering
		\includegraphics[width=\textwidth]{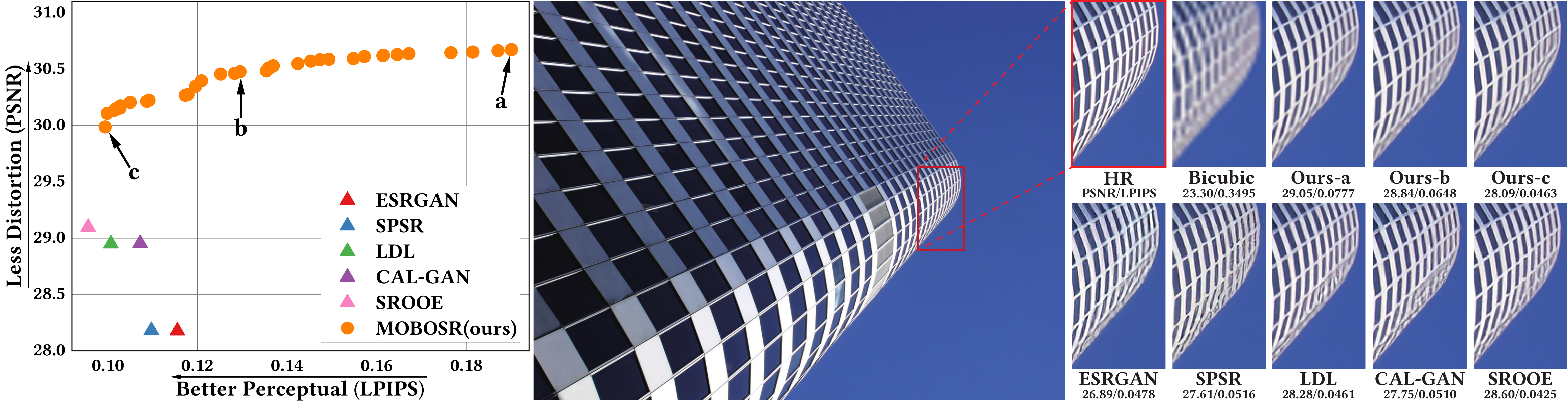}
		\caption{Left: Comparisons on the DIV2K \cite{Agustsson17} validation set, our method not only significantly surpasses others in distortion metrics (PSNR) but also holds an advantage in perceptual quality (LPIPS \cite{Zhang18}). [a, b, c] represent three different sample points on the perception-distortion Pareto frontier to illustrate the trade-off between distortion and perceptual quality. Right: From the quantitative and visual comparison of this single image, it is evident from our results, labeled as Ours-[a, b, c], that as PSNR decreases and LPIPS increases along the Pareto frontier, the images become sharper. However, this sharpness is accompanied by the introduction of noise and artifacts. In contrast to other methods, our method generates fewer artifacts and avoids excessive blurring, which can be attributed to our model achieving an optimal balance between perceptual quality and distortion.}
		\label{fig:teaser}
	\end{teaserfigure}
	
	\maketitle
	
	\section{Introduction}

Super-resolution (SR) plays a crucial role in multimedia, increasingly integrated into devices to enhance the viewing experience of low-resolution images and videos, not only improve visual quality but also reducing bandwidth and storage needs for media streaming. 

In the single-image super-resolution (SISR) field, both model architecture and computational efficiency have received considerable attention, resulting in impressive achievements. The majority of these studies report and compare results using PSNR and SSIM \cite{Zhou04}, both of which are pixel-based metrics. Thus, SR task is actually treated as a regression problem, using regression losses to train, such as L1 Normalization (Mean Absolute Error), L2 Normalization (Mean Squared Error), Charbonnier loss \cite{Lai17}, and Huber loss \cite{Laina16}. Pixel-based regression losses are favored for their simplicity and computational efficiency. However, as image downscaling results in significant loss of high-frequency details, SR poses an ill-posed problem. Pixel-based regression loss functions struggle with recovering these high-frequency details \cite{Rad19, Liang22}, often producing blurred images, despite achieving high PSNR and SSIM \cite{Zhou04} scores.

\begin{figure}[ht]
	\centering
	\includegraphics[width=0.90\linewidth]{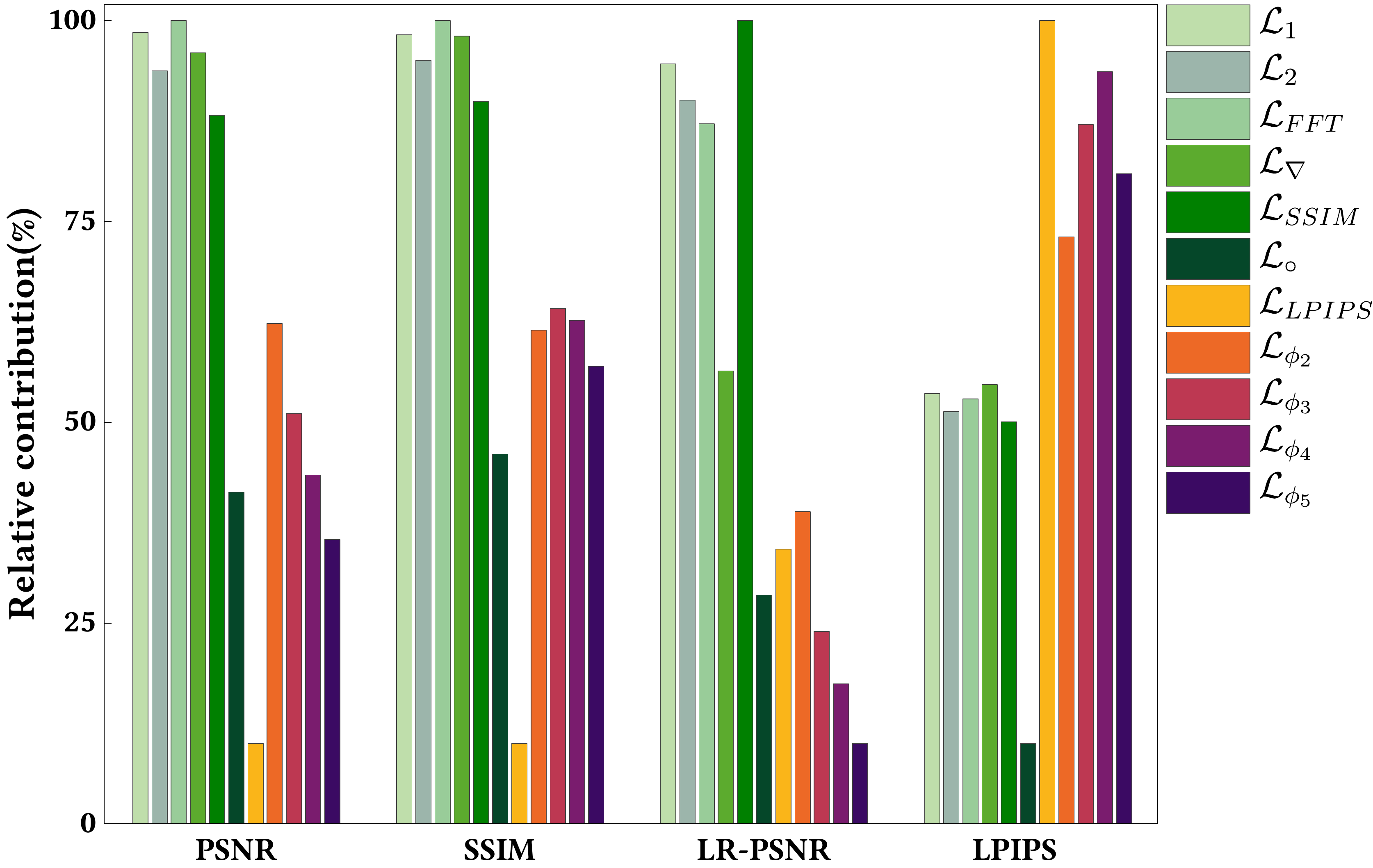}
	\caption{Relative contributions of each loss function to the final model metrics. The definitions of these loss functions are detailed in Section \ref{sec:impl_detail}. The relative contribution are calculated by training a SR model with each individual loss function and evaluating them on the DIV2K \cite{Agustsson17} validation set. Metrics are normalized, so a higher bar indicates better performance for the corresponding metric. This indicates there is an inherent conflict between perceptual and regression losses (e.g. [$\mathcal{L}_{1}$, $\mathcal{L}_{2}$, $\mathcal{L}_{FFT}$, $\mathcal{L}_{\nabla}$, $\mathcal{L}_{SSIM}$] contribute more to distortion metrics (PSNR, SSIM \cite{Zhou04}, and LR-PSNR), while [$\mathcal{L}_{LPIPS}$, $\mathcal{L}_{\phi_{2}}$, $\mathcal{L}_{\phi_{3}}$, $\mathcal{L}_{\phi_{4}}$, $\mathcal{L}_{\phi_{5}}$] contribute more to perceptual metric (LPIPS \cite{Zhang18})).}
	\label{fig:loss_compare}
\end{figure}

While PSNR or SSIM \cite{Zhou04} are convenient for computation and comparison, they do not adequately reflect the perceptual quality of images and may even have a negative correlation with human perception (see Figure \ref{fig:teaser}, `Ours-a' has higher PSNR than `Ours-c' while `Ours-a' is more blurry on the contrary). The most accurate reflection of human perceptual quality is naturally human-opinion-scores (HOS), but due to its irreproducibility and high cost, it is challenging to use for fair comparative analysis \cite{PIRM18}. Consequently, Full Reference Image Quality Assessment (FR-IQA) metrics like LPIPS \cite{Zhang18} have been developed to better align with human perception while being reproducible and computationally efficient. And to enhance the perceptual quality of SR models, researchers commonly employ perceptual loss \cite{Johnson16} or GAN \cite{Goodfellow20} for training, which, while enhancing image sharpness, may also result in the introduction of noise and artifacts (see Figure \ref{fig:teaser}, `SPSR' \cite{Ma20} holds better LPIPS \cite{Zhang18} than `Ours-b', but `SPSR' \cite{Ma20} causes much more artifacts).

Therefore, many researchers combine regression and perceptual losses to strive for an optimal balance between pixel accuracy and visual perception \cite{Vu18, Rad19, Wang19, Wang19a, Deng19, Fritsche19, Fuoli21, Liang22, Liu23, Park23}. However, as shown in Figure \ref{fig:loss_compare}, there is an inherent conflict between perceptual loss \cite{Johnson16} and regression losses, as well as between perceptual and distortion metrics \cite{Blau18}, leading to a compromise.

Several methods attempts to balance perception and distortion by manually setting the weights of different losses during training \cite{Ledig17, Sajjadi17, Park18, Wang19, Fuoli21}. Concurrently, some approaches involve training two structurally identical networks: one employs regression loss to optimize distortion metrics, while the other adopts perceptual loss \cite{Johnson16} to enhance perceptual metrics. The parameters of these networks are subsequently blended through interpolation, employing manually determined weights to finely adjust the trade-off between perception and distortion within the final model \cite{Vu18, Wang19, Wang19b}. In a novel strategy, Wang et al. \cite{Wang19a} incorporated conditional branches into the network, which are regulated by a scalar value to produce varying trade-offs. Furthermore, certain methods selectively apply different losses to different regions or frequency components of an image \cite{Wang18a, Rad19, Fritsche19, Deng19, Liang22, Liu23, Park23}. This entails utilizing perceptual loss \cite{Johnson16} or GAN \cite{Goodfellow20} for areas with intricate textures to maximize perceptual quality, while reserving pixel regression loss for regions with simpler textures to maintain fidelity. While these artworks have garnered notable success, they have not yet explored the optimal balance boundary between perception and distortion. Furthermore, they either require laborious hyperparameter tuning or impose additional computational demands during the inference phase, and can only apply a limited number of loss functions.

To intuitively illustrate the conflict between perception and distortion, we conducted a series of simple experiments to quantify the contributions of various loss functions to different IQA metrics, as depicted in Figure \ref{fig:loss_compare}. It was observed that loss functions with a higher contribution to distortion metrics inversely contributed less to perceptual metrics, and vice versa. This conflicting characteristic prompted us to consider: {\bfseries Whether Multi-Objective Optimization (MOO) could be a viable solution to this dilemma?}

MOO is an optimization strategy designed to simultaneously address multiple objectives. Due to potential contradictions among these objectives, it is often challenging to find a single solution that maximizes or minimizes all objectives concurrently. Therefore, the aim of MOO is to identify a set of solutions that offer the best compromise among the different objectives, culminating in what is known as the Pareto frontier. On this frontier, any improvement in one objective comes at the cost of compromising at least one other objective. Multi-Objective Bayesian Optimization (MOBO), a widely utilized algorithm within the MOO framework, is particularly adept at optimizing black-box objective functions that are gradient-inaccessible and costly to evaluate.

This paper introduces a novel application of MOBO to address the challenge of balancing perceptual quality and distortion in SISR models. We conceptualize the relationship between loss weights and IQA metrics as black-box objective functions to be optimized. In the training process, the performance metrics of SR models guide the optimization algorithm to refine the loss weights for the next epoch, thereby creating a feedback loop that enhances model efficacy with each iteration. This approach avoids the typical computational overhead associated with AutoML, where each optimization iteration requires a complete model retraining. It incurs no additional computational or storage costs during inference and can apply numerous loss functions during training.

In summary, the contributions of this work are threefold:
\begin{itemize}
	\item We pioneer the application of MOBO to address the challenge of balancing perception quality and distortion in SISR models.
	\item We have developed a method for dynamically adjusting loss weights during model training, thereby obviating the need for hyperparameter tuning and significantly reducing the computational resources required in comparison to AutoML hyperparameter search methods.
	\item This research expands the perceptual-distortion Pareto frontier of the SISR domain, offering new insights into the trade-offs between perceptual quality and distortion in various image restoration tasks.
\end{itemize}

	\section{Related Works}

Our work employs MOBO to address the balance between perception and distortion in SISR models. The following sections will provide an overview of SISR methods aimed at addressing this balance, followed by a concise introduction to MOO methods.

\subsection{Single Image Super Resolution}

Since Dong et al. \cite{Dong14} first applied CNN to the task of SISR, this field has been predominantly driven by deep learning advancements. SISR models have continuously improved in performance. Notably, ESPCN \cite{Shi16} has replaced interpolation with sub-pixel convolution and deferred scaling to the end of the model, significantly reducing computational demands. EDSR \cite{Lim17} has achieved enhanced performance by omitting Batch Normalization \cite{Ioffe15} layers, which, despite their effectiveness in other tasks, proved detrimental to SR tasks. Furthermore, SwinIR \cite{Liang21} has constructed SR models based on Swin Transformer \cite{Liu21}, establishing a new SOTA. These methods employ regression loss for training, achieving high PSNR and SSIM \cite{Zhou04} scores but failing to recover high-frequency details, resulting in blurred images.

Johnson et al. \cite{Johnson16} introduced perceptual loss which utilizes feature maps extracted from a pre-trained deep convolutional network (such as VGG \cite{Simonyan15}) as the input for the loss function, significantly enhancing the perceptual quality of super-resolution models. SRGAN \cite{Ledig17} pioneered the application of GAN \cite{Goodfellow20} to SR models, enabling the generation of sharp details. These two seminal works have inspired a plethora of perception-oriented SR approaches. Beyond perceptual \cite{Johnson16} and GAN \cite{Goodfellow20} losses, a plethora of other loss functions have been incorporated into SISR models. A category of these introduces additional constraints, including: frequency domain constraints \cite{Cheon19, Fuoli21, Sun22, Wang23, Liu23, Deng19}, which utilize the spectral maps generated by FFT, DCT or Wavelet transformations to compute loss; gradient constraints \cite{Cheon19, Ma20, Vu18, Umer20, Lusine22, Wang22}, which calculate loss using the gradient maps of images; and low-resolution consistency constraints \cite{Navarrete19, Umer20}, which compute loss by comparing the downsampled super-resolved images with the original low-resolution images, thereby reducing the likelihood of generating incorrect textures. The application of these losses can further enhance certain metrics of SISR models, as shown in Figure \ref{fig:loss_compare}.

\subsection{Perceptual Distortion Trade-Off}

However, while perceptual loss \cite{Johnson16} and GAN \cite{Goodfellow20} can enhance perceptual quality, they also introduce noise or artifacts and may even produce incorrect textures. To balance perceptual quality and distortion in SISR, most methods involve manually adjust the weights of different loss functions \cite{Ledig17, Sajjadi17, Park18, Wang19, Fuoli21}, necessitating extensive hyperparameter tuning to achieve superior outcomes, yet without guaranteeing optimality. Vu et al. \cite{Vu18}, along with Wang et al. \cite{Wang19, Wang19b}, have demonstrated that a balance between perception quality and distortion can be attained through network parameter interpolation. This method mitigates the need for extensive network training for each weight by limiting the process to just two networks, thereby substantially reducing resource consumption. However, it is important to note that this approach still does not guarantee optimal results. In a similar way, Wang et al. \cite{Wang19a} introduced conditional branches within the network architecture, modulated by a scalar parameter, to facilitate a spectrum of trade-offs. However, this approach introduces additional computational load during inference phase.

Rad et al. \cite{Rad19} categorized images into background, edges, and objects, applying different loss functions to each category. Fritsche et al. \cite{Fritsche19} employed separate loss functions for the high-frequency and low-frequency components of images. Liang et al. \cite{Liang22} divided images into numerous patches, classifying them into three categories—solid colors, texture-rich like trees or hair, and those with strong edge transitions—and applied different loss functions accordingly. Liu et al. \cite{Liu23} segmented images based on Spectral Bayesian Uncertainty before applying various loss functions. Park et al. \cite{Park23} train an auxiliary network using the LPIPS \cite{Zhang18} Map as supervisory signal to predict a discrete set of predefined loss weight combinations, aiming to maximum perceptual quality. These methods either fail to achieve the optimal balance, necessitate additional network training, or increase computational load during inference. Decomposing images before super-resolution and then recombining them may also potentially introduce more noise or artifacts.

\subsection{Multi-Objective Optimization}

MOO is a process that seeks to find the optimal balance among several often conflicting objectives. If the user has a clear preference for each objective, a weighted sum approach can be employed to transform it into a single-objective optimization problem. Conversely, if the user’s preferences are not well-defined, MOO techniques are necessary to determine the Pareto frontier.

In the field of MOO, the two main methodologies are multi-objective evolutionary algorithms (MOEA) and multi-objective Bayesian optimization (MOBO). MOEAs, such as NSGA-II \cite{Deb02}, SPEA2 \cite{Zitzler01}, MOEA/D \cite{Zhang07}, SMPSO \cite{Nebro09} etc., are well-regarded for their ability to generate a diverse set of Pareto-optimal solutions through population-based approaches. They excel in exploring the solution space, which is crucial for capturing the trade-offs among conflicting objectives. However, MOEAs can be computationally demanding, especially when scaling to problems with high-dimensional objective spaces or when requiring a high-resolution Pareto front.

MOBO approaches extends Bayesian Global Optimization (BGO) \cite{Jones98} from single-objective to multi-objective contexts. It typically uses Gaussian Processes (GP) for modeling the objective functions and utilizes heuristic acquisition functions, such as Expected Hypervolume Improvement (EHVI) \cite{Emmerich06, Wagner10, Daulton20, Daulton21}, to identify the subsequent data point for evaluation. MOBO is highly sample efficient and is particularly well-suited for optimizing black-box objective functions which are expensive to evaluate. Thus, MOBO is especially well-suited for implementing our proposed SISR approach that balances perceptual quality and distortion.
	\section{Method}

\subsection{Preliminary}

MOO refers to simultaneously optimizing multiple objective functions $\bm{f}(\bm{x})\subset\mathbb{R}^{M}$ over a bounded search space $\mathcal{X}\in\mathbb{R}^{d}$, aiming to find solutions that optimize all objectives. However, these objectives often conflict with each other, making it challenging to find a single solution that optimizes all objectives simultaneously. Instead, we typically obtain a compromise among multiple objectives. A common approach is to identify the {\bfseries Pareto front}. 

\begin{figure}[!t]
	\centering
	\includegraphics[width=\linewidth]{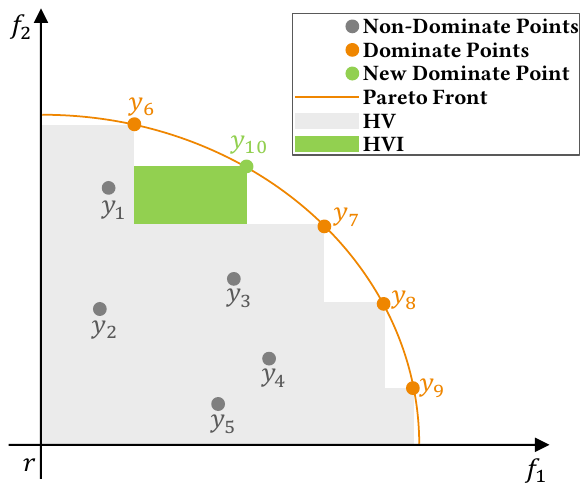}
	\caption{A toy example that demonstrates the 2-dimensional Pareto frontier, along with the HV and HVI.}
	\label{fig:ehvi}
\end{figure}

\begin{description}
	\item[\texttt{Definition 1}:] Given $\bm{f}(\bm{x})=[f_{1}(\bm{x}),\ldots, f_{M}(\bm{x})]$ and $\bm{f}(\bm{x'})=[f_{1}(\bm{x'}),\ldots, f_{M}(\bm{x'})]$, if $f_{i}(\bm{x}) \ge f_{i}(\bm{x'})$ for all $i = 1,\ldots,M$, and there exists at least one $i\in{1,\ldots,M}$ such that $f_{i}(\bm{x}) > f_{i}(\bm{x'})$, we say that $\bm{f}(\bm{x})$ {\bfseries dominates} $\bm{f}(\bm{x'})$, denoted as $\bm{f}(\bm{x}) \succ \bm{f}(\bm{x'})$.
	\item[\texttt{Definition 2}:] An $\bm{f}(\bm{x^{*}})$ is considered {\bfseries Pareto optimal} if no other solution can dominate it.
	\item[\texttt{Definition 3}:] In the space of objective functions, the set of all Pareto optimal solutions forms the {\bfseries Pareto front}: $\mathcal{P}=\{\bm{f}(\bm{x})~s.t.~\nexists\bm{x'}\in\mathcal{X}:\bm{f}(\bm{x'})\succ\bm{f}(\bm{x})\}$. 
	\item[\texttt{Definition 4}:] {\bfseries Hypervolume (HV)} quantifies the volume of the space dominated by a finite approximate Pareto front $\mathcal{P}$, with respect to a reference point $\bm{r}\in\mathbb{R}^{M}$. Specifically, it measures the M-dimensional Lebesgue measure $\lambda_{M}$ of the region that is bounded by the hyper-rectangle formed by vertices $\bm{r}$ and $\bm{y}_{i}$: $HV(\mathcal{P},\bm{r})=\lambda_{M}(\cup_{i=1}^{\left|\mathcal{P}\right|}[\bm{r},\bm{y}_{i}])$.
	\item[\texttt{Definition 5}:] {\bfseries Hypervolume Improvement (HVI)} assesses the improvement achieved by a set of points $\mathcal{Y}$ concerning a given Pareto front $\mathcal{P}$, relative to a reference point $\bm{r}$: $HVI(\mathcal{Y},\mathcal{P},\bm{r})=HV(\mathcal{P}\cup\mathcal{Y},\bm{r})-HV(\mathcal{P},\bm{r})$.
\end{description}

Figure \ref{fig:ehvi} presents an intuitive 2-dimensional toy example of these definitions. When optimizing an objective function $\bm{f}(\bm{x})$ that is expensive to evaluate and lacks gradient information, MOBO emerges as the most suitable approach. It constructs a probabilistic surrogate model representing the objective function and employs an acquisition function to determine the next sampling point. Through iterative updates of the surrogate model, MOBO can efficiently identify the global optimum with fewer evaluations. 

\subsection{Problem Formulation}

The primary goal of SISR is to employ a model, \emph{e.g.} a neural network, denoted as $G_{\theta}$ and parameterized by $\theta$, to transform a low-resolution input image $I^{LR}$, which is typically obtained through bicubic downsampling from a high-resolution image $I^{HR}$, into a super-resolved image $I^{SR}$ with the same resolution as $I^{HR}$:
\begin{equation}
	I^{SR} = G_{\theta}(I^{LR}).
\end{equation}

The goal is to make $I^{SR}$ as close as possible to $I^{HR}$, and to achieve this, we minimize a loss function:
\begin{equation}
	\mathop{\arg\min}\limits_{\theta}{\mathcal{L}(G_{\theta}(I^{LR}),I^{HR})}.
\end{equation}

However, relying solely on a single loss function may not lead to an optimal model, as different loss functions have different advantages and drawbacks. Therefore, we linearly combine multiple losses:
\begin{equation}
	\mathcal{L}=\sum_{i=1}^{N}{\omega_{i}\mathcal{L}_{i}}.
\end{equation}

The gradient of this combined loss with respect to $\theta$ is:
\begin{equation}
	\frac{\partial\mathcal{L}}{\partial\theta} =\sum_{i=1}^{N}{\omega_{i}\frac{\partial\mathcal{L}_{i}}{\partial\theta}}.
\end{equation}

Therefore, by adjusting the weights $\omega_{i}$ of each loss $\mathcal{L}_{i}$, we can control the corresponding magnitude of the loss gradient and, consequently, its impact on the IQA metrics $\bm{Q}_{\theta}=[PSNR_{\theta},\ldots,LPIPS_{\theta}]$ of SR model. There exists a deterministic relationship between $\bm{Q}_{\theta}$ and loss weights $\bm{\omega}=[\omega_{1},\ldots,\omega_{N}]$:
\begin{equation}
	\bm{Q}_{\theta}=\bm{f}(\bm{\omega}).
\end{equation}

Solving for $\bm{f}$ allows us to achieve optimal balance between perceptual quality and distortion by setting the best loss weights. However, since $\bm{f}$ is an unknown black-box function and these objectives may conflict with each other, MOBO will be employed for its estimation.

\subsection{Multi-Objective Bayesian Optimization Super-Resolution}

Initially, we randomly and uniformly sample a certain number of initial weights within the search space $\bm{\omega}\in\bm{\Omega}^{d}$, train the SR model using the sampled weights, and then compute the IQA metrics of the SR model to obtain the observed dataset $\mathcal{D}=\left\{(\bm{Q}_{i},\bm{\omega}_{i})\right\}_{i=1}^{T}$.

Subsequently, we use the observed data $\mathcal{D}$ to fit a probabilistic surrogate model. Surrogate models typically employ Gaussian processes (GP), which simulates the probability distribution of the true objective function $\bm{f}$ based on the observed data $\mathcal{D}$:
\begin{equation}
	\bm{f}\sim\mathcal{GP}(\bm{\mu},\bm{K}),
\end{equation}
where $\bm{\mu}$ is the mean function, and $\bm{K}$ is the covariance function. In MOBO, a GP is fitted for each objective.

Given the high cost of evaluating the objective function and the low cost of evaluating the surrogate model, BO utilizes acquisition functions to heuristically select the next data point for evaluating and training, thereby minimizing the number of evaluations. Acquisition functions assess the potential performance of new data point based on the distribution predicted by the surrogate model. The most commonly used acquisition function is the Expected Improvement (EI):
\begin{equation}
	\alpha_{EI}(\bm{\omega})=\mathbb{E}[\max(0,\bm{f}(\bm{\omega}^{*})-\bm{f}(\bm{\omega}))],
\end{equation}
which can achieves a good balance between exploration and exploitation. When extended to multi-objective scenarios, the Expected Hypervolume Improvement (EHVI) is used as the acquisition function:
\begin{equation}
	\alpha_{EHVI}(\bm{\omega}|\mathcal{P})=\mathbb{E}[HVI(\bm{f}(\bm{\omega})|\mathcal{P}))].
\end{equation}
Through the acquisition function, we iteratively select the next most valuable point for evaluation and update the surrogate model:
\begin{equation}
	\bm{\omega}^{*}=\mathop{\arg\max}\limits_{\bm{\omega}\in\bm{\Omega}^{d}}{\alpha(\bm{\omega})}.
\end{equation}

\begin{algorithm}[!t]
	\caption{MOBOSR Pseudocode}
	\label{alg}
	\SetAlgoLined
	\KwData{Pretrain epochs $T_{init}$, total epochs $T$, $i \leftarrow 1$.}
	\KwResult{SR models $\mathcal{M}=\left \{ M_{1},\ldots,M_{N} \right \}$ on optimal distortion and perceptual Pareto frontier.}
	Define optimization objectives $\bm{Q}$, parameters $\bm{\omega}$, dataset $\mathcal{D}$, objective function $\bm{f}\sim\mathcal{GP}(\bm{\mu},\bm{K})$, acquisition function $\alpha_{EHVI}(\bm{\omega}|\mathcal{P})$, SR model $\bm{M}$\;
	\For{$i=1$ to $T_{init}$}{
		Random sample $\bm{\omega}_{i}\in\bm{\Omega}^{d}$\;
		$\bm{M}$.train($\bm{\omega}_{i}$)\;
		$\bm{Q}_{i}$=$\bm{M}$.eval()\;
		$\mathcal{D}=\mathcal{D}\cup\left\{(\bm{Q}_{i},\bm{\omega}_{i})\right\}$\;
	}
	\For{$i=1$ to $T$}{
		Fit $\bm{f}\sim\mathcal{GP}(\bm{\mu},\bm{K})$ using $\mathcal{D}$\;
		$\bm{\omega}_{i}=\mathop{\arg\max}\limits_{\bm{\omega}\in\bm{\Omega}^{d}}{\alpha_{EHVI}(\bm{\omega}|\mathcal{P})}$\;
		$\bm{M}$.train($\bm{\omega}_{i}$)\;
		$\bm{Q}_{i}$=$\bm{M}$.eval()\;
		$\mathcal{D}=\mathcal{D}\cup\left\{(\bm{Q}_{i},\bm{\omega}_{i})\right\}$\;
	}
	Obtain $\mathcal{M}$ through compute Pareto frontier from $\mathcal{D}$.
\end{algorithm}

By repeating these steps, BO can achieve the global optimum of the black-box objective function with a minimal number of evaluation steps. The pseudocode for the overall optimization process is outlined as Algorithm \ref{alg}.

	\begin{table*}
	\caption{Comparison of MOBOSR with other artworks on 7 datasets. The best and second-best results are highlighted in \textbf{bold} and \underline{underline}, respectively. The symbols $\uparrow$ and $\downarrow$ indicate that higher or lower values of the metric are preferable. To the best of our knowledge, for fair comparisons, all publicly available methods~(utilize RRDB \cite{Wang19} as the backbone, and aim to address the balance between perceptual quality and distortion) are selected. This includes SPSR \cite{Ma20}, RRDB+LDL \cite{Liang22}, CAL-GAN \cite{Park23a} and SROOE \cite{Park23}. In this table, we use the data point labeled as Our-c in Figure \ref{fig:teaser} for comparison. Except for SROOE \cite{Park23}, our proposed MOBOSR consistently outperforms existing methods by a large margin on all metrics, particularly demonstrating a significant advantage in LR-PSNR with over 5dB improvements. Compared to SROOE \cite{Park23}, we achieve superior performance on all distortion metrics with similar LPIPS \cite{Zhang18}, even though we only train on the smaller DIV2K \cite{Agustsson17} training set. Quantitative comparisons of Ours-[a,b] and details on metric evaluation are available in the supplementary material.}
	\label{tab:compare_2_sota}
	\begin{tabular*}{\textwidth}{@{\extracolsep{\fill}}c|c|c|ccccccc}
		\toprule
		Metrics & Methods & Train Datasets & Set5 & Set14 & DIV2K & BSD100 & Urban100 & General100 & Manga109 \\
		\midrule
		\multirow{6}*{PSNR$\uparrow$}
		& ESRGAN \cite{Wang19} & DF2K-OST & 30.4618 & 26.2839 & 28.1778 & 25.2892 & 24.3617 & 29.4593 & 28.5041 \\
		& SPSR \cite{Ma20} & DIV2K & 30.3871 & 26.6501 & 28.1824 & 25.4949 & 24.8063 & 29.4794 & 28.6102 \\
		& RRDB+LDL \cite{Liang22} & DIV2K & 31.0007 & 27.2064 & 28.9510 & 26.0988 & 25.4781 & 30.1974 & 29.4111 \\
		& CAL-GAN \cite{Park23a} & DIV2K & 31.0475 & \underline{27.3272} & 28.9549 & \underline{26.2581} & 25.2908 & 30.0742 & 29.1665 \\
		& SROOE \cite{Park23} & DF2K & \underline{31.2455} & 27.2561 & \underline{29.0990} & 26.1715 & \underline{25.8452} & \underline{30.4723} & \underline{29.9017} \\
		& Ours-c & DIV2K & \textbf{31.8272} & \textbf{28.1766} & \textbf{29.9858} & \textbf{27.0494} & \textbf{26.0764} & \textbf{31.1164} & \textbf{30.2763} \\
		\hline
		\multirow{6}*{SSIM$\uparrow$}
		& ESRGAN \cite{Wang19} & DF2K-OST & 0.8518  & 0.6982  & 0.7761  & 0.6496  & 0.7341  & 0.8102  & 0.8604  \\
		& SPSR \cite{Ma20} & DIV2K & 0.8432  & 0.7133  & 0.7720  & 0.6571  & 0.7472  & 0.8095  & 0.8591  \\
		& RRDB+LDL \cite{Liang22} & DIV2K & 0.8610 & 0.7343  & 0.7952  & 0.6811  & 0.7670  & 0.8278 & 0.8746  \\
		& CAL-GAN \cite{Park23a} & DIV2K & 0.8552 & \underline{0.7353}  & 0.7897  & 0.6789  & 0.7623  & 0.8262  & 0.8676  \\
		& SROOE \cite{Park23} & DF2K & \underline{0.8651} & 0.7304 & \underline{0.7980} & \underline{0.6866} & \underline{0.7764} & \underline{0.8332} & \underline{0.8786} \\
		& Ours-c & DIV2K & \textbf{0.8804}  & \textbf{0.7615}  & \textbf{0.8203}  & \textbf{0.7109}  & \textbf{0.7812}  & \textbf{0.8495}  & \textbf{0.8938}  \\
		\hline
		\multirow{6}*{LR-PSNR$\uparrow$}
		& ESRGAN \cite{Wang19} & DF2K-OST & 46.7348 & 43.8433 & 45.9012 & 43.8190 & 42.9339 & 45.4220 & 43.9667 \\
		& SPSR \cite{Ma20} & DIV2K & 46.3607 & 43.6201 & 44.8529 & 42.6756 & 42.6679 & 44.6786 & 44.3872 \\
		& RRDB+LDL \cite{Liang22} & DIV2K & 48.5067 & 46.2893 & 47.9757 & 45.1571 & 46.5827 & 48.0079 & 47.8923 \\
		& CAL-GAN \cite{Park23a} & DIV2K & 42.4327 & 41.5963 & 42.8611 & 41.0666 & 41.6069 & 43.4227 & 42.8636 \\
		& SROOE \cite{Park23} & DF2K & \underline{53.1781} & \underline{51.0679} & \underline{53.5488} & \underline{51.2347} & \underline{50.6700} & \underline{52.9797} & \underline{51.7820} \\
		& Ours-c & DIV2K & \textbf{54.3372} & \textbf{53.3344} & \textbf{55.2161} & \textbf{53.3618} & \textbf{52.9401} & \textbf{54.5283} & \textbf{53.4195} \\
		\hline
		\multirow{6}*{LPIPS$\downarrow$}
		& ESRGAN \cite{Wang19} & DF2K-OST & 0.0750  & 0.1341  & 0.1155  & 0.1617  & 0.1228  & 0.0876  & 0.0647  \\
		& SPSR \cite{Ma20} & DIV2K & 0.0616  & 0.1313  & 0.1097  & 0.1629  & 0.1186  & 0.0866  & 0.0662  \\
		& RRDB+LDL \cite{Liang22} & DIV2K & 0.0637  & 0.1309  & 0.1007  & 0.1635  & \underline{0.1097}  & 0.0794  & \underline{0.0546}  \\
		& CAL-GAN \cite{Park23a} & DIV2K & 0.0687  & 0.1320  & 0.1072  & 0.1696  & 0.1171  & 0.0894  & 0.0688  \\
		& SROOE \cite{Park23} & DF2K & \textbf{0.0603} & \textbf{0.1131} & \textbf{0.0956} & \underline{0.1514} & \textbf{0.1067} & \textbf{0.0758} & \textbf{0.0511} \\
		& Ours-c & DIV2K & \underline{0.0607}  & \underline{0.1240}  & \underline{0.0994}  & \textbf{0.1508}  & 0.1154  & \underline{0.0776}  & 0.0576 \\
		\bottomrule
	\end{tabular*}
\end{table*}

\begin{figure*}[!ht]
	\centering
	\includegraphics[width=\textwidth]{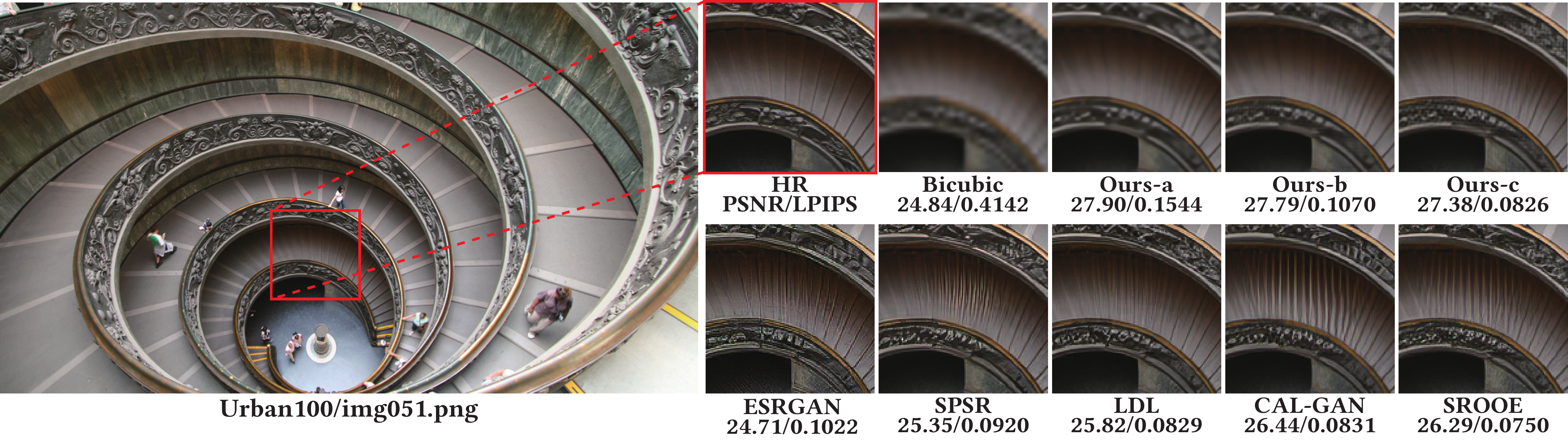}
	\includegraphics[width=\textwidth]{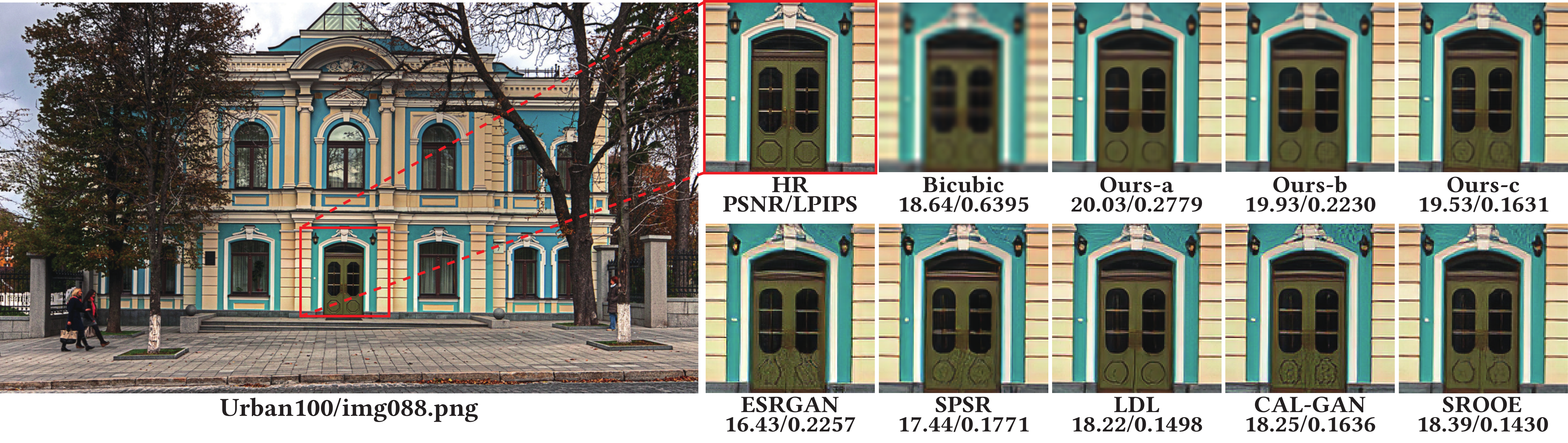}
	\caption{Visual comparison of three sampled points on the Pareto frontier obtained through our method (as defined in Figure \ref{fig:teaser}), alongside other artworks, on the Urban100 \cite{Huang15} dataset. More visual results are presented in the supplementary material.}
	\label{fig:visual_compare}
\end{figure*}

\section{Experiments}

\subsection{Implementation Details}
\label{sec:impl_detail}

\paragraph{Model Architecture}
The ESRGAN \cite{Wang19} architecture was employed as the SR model to evaluate the approach, due to its excellent performance and popularity in recent perception-related SR research. The widespread adoption of ESRGAN \cite{Wang19} and its proposed RRDB \cite{Wang19} backbone ensures a comprehensive and equitable comparison of our methodology.

\paragraph{Loss Functions}
The loss functions we employed include:
\begin{itemize}
	\item L1 Loss:
	\begin{equation}
		\mathcal{L}_{1}=\left\|I^{SR}-I^{HR}\right\|_{1}.
	\end{equation}
	\item L2 Loss:
	\begin{equation}
		\mathcal{L}_{2}=\left\|I^{SR}-I^{HR}\right\|_{2}^{2}.
	\end{equation}
	\item FFT Loss:
	\begin{equation}
		\mathcal{L}_{FFT}=\left\|FFT(I^{SR})-FFT(I^{HR})\right\|_{1}.
	\end{equation}
	\item Gradient Loss:
	\begin{equation}
		\mathcal{L}_{\nabla}=\left\|\nabla_{h}I^{SR}-\nabla_{h}I^{HR}\right\|_{1}+\left\|\nabla_{v}I^{SR}-\nabla_{v}I^{HR}\right\|_{1},
	\end{equation}
	where $\nabla_{h}$ and $\nabla_{v}$ are the Sobel gradient operators in the horizontal and vertical directions.
	\item Perceptual Loss \cite{Johnson16}:
	\begin{equation}
		\mathcal{L}_{\phi_{j}}=\left\|\phi_{j}(I^{SR})-\phi_{j}(I^{HR})\right\|_{1},j\in\{2,3,4,5\},
	\end{equation}
	where $\phi_{j}$ is the feature map output by the VGG \cite{Simonyan15} network’s $j$th layer.
	\item Cycle Consistency Loss:
	\begin{equation}
		\mathcal{L}_{\circ}=\left\|\Downarrow^{3}(I^{SR})-I^{LR}\right\|_{1},
	\end{equation}
	where $\Downarrow^{3}$ is the bicubic downsampling operator.
	\item GAN Loss \cite{Goodfellow20}:
	
	The discriminator loss is:
	\begin{equation}
		\mathcal{L}_{D}=-log(D(I^{SR})),
	\end{equation}
	where $D$ represents the discriminator network. The generator loss is:
	\begin{equation}
		\mathcal{L}_{GAN}=-log(D(I^{HR}))-log(1-D(I^{SR})).
	\end{equation}
	\item SSIM \cite{Zhou04} Loss:
	\begin{equation}
		\mathcal{L}_{SSIM}=1-\frac{{(2\mu_{I^{SR}}\mu_{I^{HR}} + C_1)(2\sigma_{{I^{SR}}{I^{HR}}} + C_2)}}{{(\mu_{I^{SR}}^2 + \mu_{I^{HR}}^2 + C_1)(\sigma_{I^{SR}} + \sigma_{I^{HR}} + C_2)}},
	\end{equation}
	where $\mu_{I^{SR}}$ and $\mu_{I^{HR}}$ are the mean values of ${I^{SR}}$ and ${I^{HR}}$, $\sigma_{I^{SR}}$ and $\sigma_{I^{HR}}$ are the standard deviations of ${I^{SR}}$ and ${I^{HR}}$, $\sigma_{{I^{SR}}{I^{HR}}}$ is the covariance between ${I^{SR}}$ and ${I^{HR}}$, $C_1$ and $C_2$ are small constants to stabilize the division.
	\item LPIPS \cite{Zhang18} Loss:
	\begin{equation}
		\mathcal{L}_{LPIPS}=\sum_{l \in F} w_l \left\|\phi_l(I^{SR})-\phi_l(I^{HR}))\right\|_{2}^{2},
	\end{equation}
	where $F$ represents the set of intermediate layers in the chosen neural network, $\phi_l$ is the normalized outputs of layer $l$ for images, $\omega_l$ are the official weights determined by Zhang et al. \cite{Zhang18}.
\end{itemize}

\paragraph{Datasets}
The DIV2K \cite{Agustsson17} training set (800 images) was employed for training. For testing, we utilize a collection of datasets including Set5 \cite{Marco12}, Set14 \cite{Zeyde12}, DIV2K \cite{Agustsson17} validation set, BSD100 \cite{Martin01}, Urban100 \cite{Huang15}, General100 \cite{Dong16}, and Manga109 \cite{Matsui17}. We followed the previous work \cite{Wang19, Ma20, Liang22, Park23a} to focus only on $\times4$ super-resolution to train and evaluate the effectiveness of our SR model. Specifically, we obtain the LR images by downsampling the HR counterparts by a factor of 4 using bicubic interpolation.

\paragraph{Training Details}
We randomly sample 16 patches from each image as a batch, the patch size of LR is $32\times32$. To ensure stability in the SR model training progress and MOBO optimization process, we initially pre-train for 250 epochs using manually set loss weights from ESRGAN \cite{Wang19}, defined as:
\begin{equation}
	\mathcal{L}=\omega_{1}\mathcal{L}_{1}+\omega_{\phi_{5}}\mathcal{L}_{\phi_{5}}+\omega_{GAN}\mathcal{L}_{GAN},
\end{equation}
where $[\omega_{1},\omega_{\phi_{5}},\omega_{GAN}]=[1e-2,1,5e-3]$. Subsequently, we commence loss weight optimization using our MOBO method. The learning rate during pre-training is fixed at $1e-4$, and for optimization with MOBO, the initial learning rate is set at $5e-5$ and is halved every 250 epochs. The optimizer employed is Adam \cite{Diederik17}. Our SR model is implemented by PyTorch \cite{Paszke19} on NVIDIA RTX 3090 GPUs, and MOBO is realized through the Ax platform \cite{Bakshy18}.

\begin{table*}
	\caption{Ablation study on 7 datasets compares MOBO-optimized loss weights to ESRGAN \cite{Wang19} manually-set loss weights. Both models use the official ESRGAN \cite{Wang19} train code and are trained on the DIV2K \cite{Agustsson17} training set with the same settings, except that the loss weights for MOBOSR are optimized by MOBO. The best results are highlighted in \textbf{bold}. MOBOSR consistently outperforms ESRGAN \cite{Wang19} by a large margin in the same setting. This comparison intuitively demonstrates the effectiveness of the core idea of our work: automatic loss weight optimization.}
	\label{tab:ablation}
	\begin{tabular*}{\textwidth}{@{\extracolsep{\fill}}c|c|cccccccc}
		\toprule
		Metrics & Methods & Set5 & Set14 & DIV2K & BSD100 & Urban100 & General100 & Manga109 \\
		\midrule
		\multirow{2}*{PSNR$\uparrow$}
		& ESRGAN \cite{Wang19} & 29.8023 & 25.5164 & 27.6994 & 25.2100 & 23.7431 & 28.8618 & 27.3669 \\
		& MOBOSR (Ours) & \textbf{30.6926} & \textbf{26.8148} & \textbf{28.5089} & \textbf{26.0402} & \textbf{24.4895} & \textbf{29.6640} & \textbf{28.2091} \\
		\hline
		\multirow{2}*{SSIM$\uparrow$}
		& ESRGAN \cite{Wang19} & 0.8456 & 0.6855 & 0.7610 & 0.6463 & 0.7116 & 0.7979 & 0.8436 \\
		& MOBOSR (Ours) & \textbf{0.8607} & \textbf{0.7234} & \textbf{0.7834} & \textbf{0.6781} & \textbf{0.7357} & \textbf{0.8209} & \textbf{0.8621} \\
		\hline
		\multirow{2}*{LR-PSNR$\uparrow$}
		& ESRGAN \cite{Wang19} & 42.5573 & 38.2986 & 41.2244 & 40.3034 & 37.6727 & 41.6356 & 40.0483 \\
		& MOBOSR (Ours) & \textbf{44.6401} & \textbf{43.0412} & \textbf{44.6847} & \textbf{43.5394} & \textbf{41.8108} & \textbf{44.6812} & \textbf{43.1860} \\
		\hline
		\multirow{2}*{LPIPS$\downarrow$}
		& ESRGAN \cite{Wang19} & \textbf{0.0742} & 0.1689 & 0.1193 & 0.1751 & 0.1379 & 0.0943 & 0.0748 \\
		& MOBOSR (Ours) & 0.0745 & \textbf{0.1359} & \textbf{0.1145} & \textbf{0.1719} & \textbf{0.1324} & \textbf{0.0894} & \textbf{0.0675} \\
		\bottomrule
	\end{tabular*}
\end{table*}

\subsection{Quantitative Results}

Methods utilize RRDB \cite{Wang19} as the backbone, and aim to address the balance between perceptual quality and distortion are selected for fair comparisons. To this end, to the best of our knowledge, all five publicly available artworks are used for comparison: ESRGAN \cite{Wang19}, SPSR \cite{Ma20}, RRDB+LDL \cite{Liang22}, CAL-GAN \cite{Park23a} and SROOE \cite{Park23}. All of which utilize RRDB \cite{Wang19} as their backbones. It is noteworthy that SROOE \cite{Park23} is trained on the larger DF2K dataset (3450 images), which consists of the DIV2K \cite{Agustsson17} training set (800 images) and Flickr2K \cite{Timofte17} (2650 images). ESRGAN \cite{Wang19}, on the other hand, is trained on the even larger DF2K-OST dataset (13774 images), which includes DF2K (3450 images) and OST \cite{Wang18a} (10,324 images). Additionally, the SPSR \cite{Ma20} network incorporates an extra gradient branch, and SROOE \cite{Park23} introduces an auxiliary network to predict a discrete set of predefined loss weight combinations. These two methods introduce additional computational load during inference.

We incorporated all the loss functions introduced in Section \ref{sec:impl_detail} and optimized their weights using the MOBO strategy. Our optimization objectives were PSNR and LPIPS \cite{Zhang18}. The loss function is formulated as:
\begin{equation}
	\begin{split}
		\mathcal{L}=&\omega_{1}\mathcal{L}_{1}+\omega_{2}\mathcal{L}_{2}+\omega_{FFT}\mathcal{L}_{FFT}+\omega_{\nabla}\mathcal{L}_{\nabla}+ \\
		&\omega_{\phi_{2}}\mathcal{L}_{\phi_{2}}+\omega_{\phi_{3}}\mathcal{L}_{\phi_{3}}+\omega_{\phi_{4}}\mathcal{L}_{\phi_{4}}+\omega_{\phi_{5}}\mathcal{L}_{\phi_{5}}+ \\
		&\omega_{\circ}\mathcal{L}_{\circ}+\omega_{GAN}\mathcal{L}_{GAN}+\omega_{SSIM}\mathcal{L}_{SSIM}+ \\
		&\omega_{LPIPS}\mathcal{L}_{LPIPS}.
	\end{split}
\end{equation}
And the optimization formula is:
\begin{equation}
	\begin{split}
		[PSNR, LPIPS]=\bm{f}([&\omega_{1}, \omega_{2}, \omega_{FFT}, \omega_{\nabla}, \omega_{\phi_{2}}, \omega_{\phi_{3}}, \omega_{\phi_{4}}, \\
		&\omega_{\phi_{5}}, \omega_{\circ}, \omega_{GAN}, \omega_{SSIM}, \omega_{LPIPS}]).
	\end{split}
\end{equation}

As demonstrated in Table \ref{tab:compare_2_sota}, except for SROOE \cite{Park23}, our approach not only demonstrates superior performance in the full-reference perceptual metric (LPIPS \cite{Zhang18}) but also significantly surpasses other methods in distortion metrics (PSNR and SSIM \cite{Zhou04}) and consistency metric (LR-PSNR). Compared to SROOE \cite{Park23}, we achieve superior performance on all distortion metrics with similar LPIPS \cite{Zhang18}, even though we only train on the smaller DIV2K \cite{Agustsson17} training set. This indicates that our approach not only achieves excellent perceptual metrics but also mitigates the artifacts and false textures introduced by perceptual loss \cite{Johnson16} and GAN \cite{Goodfellow20}, thereby ensuring consistency with the LR source and the pixel accuracy with the HR image. In summary, our method consistently outperforms others in both distortion and perceptual quality, demonstrating that the integration of MOBO facilitates the attainment of the Pareto frontier (optimal balance boundary) for perception quality and distortion.

Furthermore, prior to our work, manually setting such a multitude of loss weights is impractical, which consequently limits the incorporation of additional loss functions into the SISR model. While our method can efficiently and autonomously determine the optimal weights for loss functions, without introducing any additional computations during the inference and significantly extending the training duration (time consuming about SR model and MOBO during training are detailed in the supplementary material).

\subsection{Qualitative Results}

Visual comparisons of Ours-[a,b,c] (as defined in Figure \ref{fig:teaser}), as depicted in Figure \ref{fig:visual_compare}, clearly reveal the inherent contradiction between perception quality and distortion. This is evidenced by the gradual deterioration in PSNR and the corresponding improvement in LPIPS \cite{Zhang18} on the Pareto frontier. A higher PSNR typically results in a more blurred image, albeit with reduced noise and artifacts. Conversely, a better LPIPS \cite{Zhang18} score indicates a sharper image, which, however, tends to introduce noise and artifacts. Compared to other methods, our approach generates fewer noise and artifacts without excessively blurring the image. Additionally, it exhibits more accurate reconstruction in certain areas, better preserving the original structure of the objects in image.

\subsection{Ablation Study}

To demonstrate the effectiveness of using MOBO to optimize loss weights compared to manually set loss weights, we conduct a comparative experiment. Our setup is identical to that of ESRGAN \cite{Wang19}, with the exception that the loss weights are optimized by MOBO. Our optimization formula is expressed as:
\begin{equation}
	[PSNR, LPIPS]=\bm{f}([\omega_{1},\omega_{\phi_{5}},\omega_{GAN}]).
\end{equation}
The loss weights for ESRGAN \cite{Wang19} are adopted from the weights utilized by the authors in the original publication: $[\omega_{1},\omega_{\phi_{5}},\omega_{GAN}]=[1e-2,1,5e-3]$. As demonstrated in Table \ref{tab:ablation}, our approach significantly surpasses ESRGAN \cite{Wang19} across all metrics, which is retrained on the DIV2K \cite{Agustsson17} training set using the official released codes. This experiment clearly demonstrates the substantial advantages of integrating MOBO for automatically optimizing loss weights in the SR model over the manual tuning of loss weights.

From the comparison between `MOBOSR (Ours)' in Table \ref{tab:ablation} and `Ours-c' in Table \ref{tab:compare_2_sota}, it is evident that incorporating a greater variety of loss functions significantly benefits the SISR model, provided that the weights for these losses are optimally set.

	\section{Conclusion}
	
	In this research, we propose the Multi-Objective Bayesian Optimization Super-Resolution (MOBOSR). To effectively address the balance between perceptual quality and distortion, we introduce multi-objective Bayesian optimization into the single-image super-resolution model, dynamically adjusting the weights of various loss functions during training. Our method is rigorously validated through comprehensive experiments, demonstrating its effectiveness and versatility in both distortion and perceptual quality. Given its potential to be applied to nearly all domains of image restoration and enhancement, we believe our method can offer valuable insights and contributions to the community.
	
	\begin{acks}
		This work was supported in part by the Key Laboratory of Smart Earth under Grant KF2023YB01-13, and in part by the National Natural Science Foundation of China (NSFC) under Grant 62176100. The computation is completed in the HPC Platform of Huazhong University of Science and Technology.
	\end{acks}
	
	\clearpage
	\bibliographystyle{ACM-Reference-Format}
	\bibliography{refs}


\begin{thebibliography}{59}

%%% the Web address.

\ifx \showCODEN    \undefined \def \showCODEN     #1{\unskip}     \fi
\ifx \showDOI      \undefined \def \showDOI       #1{#1}\fi
\ifx \showISBNx    \undefined \def \showISBNx     #1{\unskip}     \fi
\ifx \showISBNxiii \undefined \def \showISBNxiii  #1{\unskip}     \fi
\ifx \showISSN     \undefined \def \showISSN      #1{\unskip}     \fi
\ifx \showLCCN     \undefined \def \showLCCN      #1{\unskip}     \fi
\ifx \shownote     \undefined \def \shownote      #1{#1}          \fi
\ifx \showarticletitle \undefined \def \showarticletitle #1{#1}   \fi
\ifx \showURL      \undefined \def \showURL       {\relax}        \fi
% invisible to TeX
\providecommand\bibfield[2]{#2}
\providecommand\bibinfo[2]{#2}
\providecommand\natexlab[1]{#1}
\providecommand\showeprint[2][]{arXiv:#2}

\bibitem[Agustsson and Timofte(2017)]%
        {Agustsson17}
\bibfield{author}{\bibinfo{person}{Eirikur Agustsson} {and}
  \bibinfo{person}{Radu Timofte}.} \bibinfo{year}{2017}\natexlab{}.
\newblock \showarticletitle{NTIRE 2017 Challenge on Single Image
  Super-Resolution: Dataset and Study}. In
  \bibinfo{booktitle}{\emph{Proceedings of the IEEE/CVF Conference on Computer
  Vision and Pattern Recognition Workshops (CVPRW)}}.
  \bibinfo{pages}{1122--1131}.
\newblock


\bibitem[Bakshy et~al\mbox{.}(2018)]%
        {Bakshy18}
\bibfield{author}{\bibinfo{person}{Eytan Bakshy}, \bibinfo{person}{Lili
  Dworkin}, \bibinfo{person}{Brian Karrer}, \bibinfo{person}{Konstantin
  Kashin}, \bibinfo{person}{Ben Letham}, \bibinfo{person}{Ashwin Murthy}, {and}
  \bibinfo{person}{Shaun Singh}.} \bibinfo{year}{2018}\natexlab{}.
\newblock \showarticletitle{AE: A domain-agnostic platform for adaptive
  experimentation}. In \bibinfo{booktitle}{\emph{Proceedings of the Advances in
  Neural Information Processing Systems Workshops (NeurIPSW)}}.
\newblock


\bibitem[Bevilacqua et~al\mbox{.}(2012)]%
        {Marco12}
\bibfield{author}{\bibinfo{person}{Marco Bevilacqua}, \bibinfo{person}{Aline
  Roumy}, \bibinfo{person}{Christine~M. Guillemot}, {and}
  \bibinfo{person}{Marie-Line Alberi-Morel}.} \bibinfo{year}{2012}\natexlab{}.
\newblock \showarticletitle{Low-Complexity Single-Image Super-Resolution based
  on Nonnegative Neighbor Embedding}. In \bibinfo{booktitle}{\emph{Proceedings
  of the British Machine Vision Conference (BMVC)}}.
\newblock


\bibitem[Blau et~al\mbox{.}(2019)]%
        {PIRM18}
\bibfield{author}{\bibinfo{person}{Yochai Blau}, \bibinfo{person}{Roey
  Mechrez}, \bibinfo{person}{Radu Timofte}, \bibinfo{person}{Tomer Michaeli},
  {and} \bibinfo{person}{Lihi Zelnik-Manor}.} \bibinfo{year}{2019}\natexlab{}.
\newblock \showarticletitle{The 2018 PIRM Challenge on Perceptual Image
  Super-Resolution}. In \bibinfo{booktitle}{\emph{Proceedings of the European
  Conference on Computer Vision Workshops (ECCVW)}}. \bibinfo{pages}{334--355}.
\newblock


\bibitem[Blau and Michaeli(2018)]%
        {Blau18}
\bibfield{author}{\bibinfo{person}{Yochai Blau} {and} \bibinfo{person}{Tomer
  Michaeli}.} \bibinfo{year}{2018}\natexlab{}.
\newblock \showarticletitle{The Perception-Distortion Tradeoff}. In
  \bibinfo{booktitle}{\emph{Proceedings of the IEEE/CVF Conference on Computer
  Vision and Pattern Recognition (CVPR)}}. \bibinfo{pages}{6228--6237}.
\newblock


\bibitem[Cheon et~al\mbox{.}(2019)]%
        {Cheon19}
\bibfield{author}{\bibinfo{person}{Manri Cheon}, \bibinfo{person}{Jun-Hyuk
  Kim}, \bibinfo{person}{Jun-Ho Choi}, {and} \bibinfo{person}{Jong-Seok Lee}.}
  \bibinfo{year}{2019}\natexlab{}.
\newblock \showarticletitle{Generative Adversarial Network-Based Image
  Super-Resolution Using Perceptual Content Losses}. In
  \bibinfo{booktitle}{\emph{Proceedings of the European Conference on Computer
  Vision Workshops (ECCVW)}}. \bibinfo{pages}{51--62}.
\newblock


\bibitem[Daulton et~al\mbox{.}(2020)]%
        {Daulton20}
\bibfield{author}{\bibinfo{person}{Samuel Daulton}, \bibinfo{person}{Maximilian
  Balandat}, {and} \bibinfo{person}{Eytan Bakshy}.}
  \bibinfo{year}{2020}\natexlab{}.
\newblock \showarticletitle{Differentiable Expected Hypervolume Improvement for
  Parallel Multi-Objective Bayesian Optimization}. In
  \bibinfo{booktitle}{\emph{Proceedings of the Advances in Neural Information
  Processing Systems (NeurIPS)}}, Vol.~\bibinfo{volume}{33}.
  \bibinfo{pages}{9851--9864}.
\newblock


\bibitem[Daulton et~al\mbox{.}(2021)]%
        {Daulton21}
\bibfield{author}{\bibinfo{person}{Samuel Daulton}, \bibinfo{person}{Maximilian
  Balandat}, {and} \bibinfo{person}{Eytan Bakshy}.}
  \bibinfo{year}{2021}\natexlab{}.
\newblock \showarticletitle{Parallel Bayesian Optimization of Multiple Noisy
  Objectives with Expected Hypervolume Improvement}. In
  \bibinfo{booktitle}{\emph{Proceedings of the Advances in Neural Information
  Processing Systems (NeurIPS)}}, Vol.~\bibinfo{volume}{34}.
  \bibinfo{pages}{2187--2200}.
\newblock


\bibitem[Deb et~al\mbox{.}(2002)]%
        {Deb02}
\bibfield{author}{\bibinfo{person}{Kalyanmoy Deb}, \bibinfo{person}{Amrit
  Pratap}, \bibinfo{person}{Sameer Agarwal}, {and} \bibinfo{person}{T.
  Meyarivan}.} \bibinfo{year}{2002}\natexlab{}.
\newblock \showarticletitle{A Fast and Elitist Multiobjective Genetic
  Algorithm: NSGA-II}.
\newblock \bibinfo{journal}{\emph{IEEE Transactions on Evolutionary
  Computation}} \bibinfo{volume}{6}, \bibinfo{number}{2}
  (\bibinfo{year}{2002}), \bibinfo{pages}{182--197}.
\newblock


\bibitem[Deng et~al\mbox{.}(2019)]%
        {Deng19}
\bibfield{author}{\bibinfo{person}{Xin Deng}, \bibinfo{person}{Ren Yang},
  \bibinfo{person}{Mai Xu}, {and} \bibinfo{person}{Pier~Luigi Dragotti}.}
  \bibinfo{year}{2019}\natexlab{}.
\newblock \showarticletitle{Wavelet Domain Style Transfer for an Effective
  Perception-Distortion Tradeoff in Single Image Super-Resolution}. In
  \bibinfo{booktitle}{\emph{Proceedings of the IEEE/CVF International
  Conference on Computer Vision (ICCV)}}. \bibinfo{pages}{3076--3085}.
\newblock


\bibitem[Dong et~al\mbox{.}(2014)]%
        {Dong14}
\bibfield{author}{\bibinfo{person}{Chao Dong}, \bibinfo{person}{Chen~Change
  Loy}, \bibinfo{person}{Kaiming He}, {and} \bibinfo{person}{Xiaoou Tang}.}
  \bibinfo{year}{2014}\natexlab{}.
\newblock \showarticletitle{Learning a Deep Convolutional Network for Image
  Super-Resolution}. In \bibinfo{booktitle}{\emph{Proceedings of the European
  Conference on Computer Vision (ECCV)}}. \bibinfo{pages}{184--199}.
\newblock


\bibitem[Dong et~al\mbox{.}(2016)]%
        {Dong16}
\bibfield{author}{\bibinfo{person}{Chao Dong}, \bibinfo{person}{Chen~Change
  Loy}, {and} \bibinfo{person}{Xiaoou Tang}.} \bibinfo{year}{2016}\natexlab{}.
\newblock \showarticletitle{Accelerating the Super-Resolution Convolutional
  Neural Network}. In \bibinfo{booktitle}{\emph{Proceedings of the European
  Conference on Computer Vision (ECCV)}}. \bibinfo{pages}{391--407}.
\newblock


\bibitem[Emmerich et~al\mbox{.}(2006)]%
        {Emmerich06}
\bibfield{author}{\bibinfo{person}{Michael T.~M. Emmerich},
  \bibinfo{person}{Kyriakos~C. Giannakoglou}, {and} \bibinfo{person}{Boris
  Naujoks}.} \bibinfo{year}{2006}\natexlab{}.
\newblock \showarticletitle{Single- and Multiobjective Evolutionary
  Optimization Assisted by Gaussian Random Field Metamodels}.
\newblock \bibinfo{journal}{\emph{IEEE Transactions on Evolutionary
  Computation}} \bibinfo{volume}{10}, \bibinfo{number}{4}
  (\bibinfo{year}{2006}), \bibinfo{pages}{421--439}.
\newblock


\bibitem[Fritsche et~al\mbox{.}(2019)]%
        {Fritsche19}
\bibfield{author}{\bibinfo{person}{Manuel Fritsche}, \bibinfo{person}{Shuhang
  Gu}, {and} \bibinfo{person}{Radu Timofte}.} \bibinfo{year}{2019}\natexlab{}.
\newblock \showarticletitle{Frequency Separation for Real-World
  Super-Resolution}. In \bibinfo{booktitle}{\emph{Proceedings of the IEEE/CVF
  International Conference on Computer Vision Workshops (ICCVW)}}.
  \bibinfo{pages}{3599--3608}.
\newblock


\bibitem[Fuoli et~al\mbox{.}(2021)]%
        {Fuoli21}
\bibfield{author}{\bibinfo{person}{Dario Fuoli}, \bibinfo{person}{Luc
  Van~Gool}, {and} \bibinfo{person}{Radu Timofte}.}
  \bibinfo{year}{2021}\natexlab{}.
\newblock \showarticletitle{Fourier Space Losses for Efficient Perceptual Image
  Super-Resolution}. In \bibinfo{booktitle}{\emph{Proceedings of the IEEE/CVF
  International Conference on Computer Vision (ICCV)}}.
  \bibinfo{pages}{2340--2349}.
\newblock


\bibitem[Goodfellow et~al\mbox{.}(2020)]%
        {Goodfellow20}
\bibfield{author}{\bibinfo{person}{Ian Goodfellow}, \bibinfo{person}{Jean
  Pouget-Abadie}, \bibinfo{person}{Mehdi Mirza}, \bibinfo{person}{Bing Xu},
  \bibinfo{person}{David Warde-Farley}, \bibinfo{person}{Sherjil Ozair},
  \bibinfo{person}{Aaron Courville}, {and} \bibinfo{person}{Yoshua Bengio}.}
  \bibinfo{year}{2020}\natexlab{}.
\newblock \showarticletitle{Generative adversarial networks}.
\newblock \bibinfo{journal}{\emph{Commun. ACM}} \bibinfo{volume}{63},
  \bibinfo{number}{11} (\bibinfo{date}{oct} \bibinfo{year}{2020}),
  \bibinfo{pages}{139–144}.
\newblock


\bibitem[Huang et~al\mbox{.}(2015)]%
        {Huang15}
\bibfield{author}{\bibinfo{person}{Jia-Bin Huang}, \bibinfo{person}{Abhishek
  Singh}, {and} \bibinfo{person}{Narendra Ahuja}.}
  \bibinfo{year}{2015}\natexlab{}.
\newblock \showarticletitle{Single Image Super-Resolution From Transformed
  Self-Exemplars}. In \bibinfo{booktitle}{\emph{Proceedings of the IEEE/CVF
  Conference on Computer Vision and Pattern Recognition (CVPR)}}.
  \bibinfo{pages}{5197--5206}.
\newblock


\bibitem[Ioffe and Szegedy(2015)]%
        {Ioffe15}
\bibfield{author}{\bibinfo{person}{Sergey Ioffe} {and}
  \bibinfo{person}{Christian Szegedy}.} \bibinfo{year}{2015}\natexlab{}.
\newblock \showarticletitle{Batch Normalization: Accelerating Deep Network
  Training by Reducing Internal Covariate Shift}. In
  \bibinfo{booktitle}{\emph{Proceedings of the International Conference on
  Machine Learning (ICML)}}, Vol.~\bibinfo{volume}{37}.
  \bibinfo{pages}{448–456}.
\newblock


\bibitem[Johnson et~al\mbox{.}(2016)]%
        {Johnson16}
\bibfield{author}{\bibinfo{person}{Justin Johnson}, \bibinfo{person}{Alexandre
  Alahi}, {and} \bibinfo{person}{Li Fei-Fei}.} \bibinfo{year}{2016}\natexlab{}.
\newblock \showarticletitle{Perceptual Losses for Real-Time Style Transfer and
  Super-Resolution}. In \bibinfo{booktitle}{\emph{Proceedings of the European
  Conference on Computer Vision (ECCV)}}. \bibinfo{pages}{694--711}.
\newblock


\bibitem[Jones et~al\mbox{.}(1998)]%
        {Jones98}
\bibfield{author}{\bibinfo{person}{Donald~R. Jones}, \bibinfo{person}{Matthias
  Schonlau}, {and} \bibinfo{person}{William~J. Welch}.}
  \bibinfo{year}{1998}\natexlab{}.
\newblock \showarticletitle{Efficient Global Optimization of Expensive
  Black-Box Functions}.
\newblock \bibinfo{journal}{\emph{Journal of Global Optimization}}
  \bibinfo{volume}{13}, \bibinfo{number}{4} (\bibinfo{date}{01 Dec}
  \bibinfo{year}{1998}), \bibinfo{pages}{455--492}.
\newblock


\bibitem[Kingma and Ba(2017)]%
        {Diederik17}
\bibfield{author}{\bibinfo{person}{Diederik~P. Kingma} {and}
  \bibinfo{person}{Jimmy Ba}.} \bibinfo{year}{2017}\natexlab{}.
\newblock \showarticletitle{Adam: A Method for Stochastic Optimization}. In
  \bibinfo{booktitle}{\emph{Proceedings of the International Conference on
  Learning Representations (ICLR)}}.
\newblock


\bibitem[Lai et~al\mbox{.}(2017)]%
        {Lai17}
\bibfield{author}{\bibinfo{person}{Wei-Sheng Lai}, \bibinfo{person}{Jia-Bin
  Huang}, \bibinfo{person}{Narendra Ahuja}, {and} \bibinfo{person}{Ming-Hsuan
  Yang}.} \bibinfo{year}{2017}\natexlab{}.
\newblock \showarticletitle{Deep Laplacian Pyramid Networks for Fast and
  Accurate Super-Resolution}. In \bibinfo{booktitle}{\emph{Proceedings of the
  IEEE/CVF Conference on Computer Vision and Pattern Recognition (CVPR)}}.
  \bibinfo{pages}{5835--5843}.
\newblock


\bibitem[Laina et~al\mbox{.}(2016)]%
        {Laina16}
\bibfield{author}{\bibinfo{person}{Iro Laina}, \bibinfo{person}{Christian
  Rupprecht}, \bibinfo{person}{Vasileios Belagiannis},
  \bibinfo{person}{Federico Tombari}, {and} \bibinfo{person}{Nassir Navab}.}
  \bibinfo{year}{2016}\natexlab{}.
\newblock \showarticletitle{Deeper Depth Prediction with Fully Convolutional
  Residual Networks}. In \bibinfo{booktitle}{\emph{Proceedings of the
  International Conference on 3D Vision (3DV)}}. \bibinfo{pages}{239--248}.
\newblock


\bibitem[Ledig et~al\mbox{.}(2017)]%
        {Ledig17}
\bibfield{author}{\bibinfo{person}{Christian Ledig}, \bibinfo{person}{Lucas
  Theis}, \bibinfo{person}{Ferenc Huszár}, \bibinfo{person}{Jose Caballero},
  \bibinfo{person}{Andrew Cunningham}, \bibinfo{person}{Alejandro Acosta},
  \bibinfo{person}{Andrew Aitken}, \bibinfo{person}{Alykhan Tejani},
  {et~al\mbox{.}}} \bibinfo{year}{2017}\natexlab{}.
\newblock \showarticletitle{Photo-Realistic Single Image Super-Resolution Using
  a Generative Adversarial Network}. In \bibinfo{booktitle}{\emph{Proceedings
  of the IEEE/CVF Conference on Computer Vision and Pattern Recognition
  (CVPR)}}. \bibinfo{pages}{105--114}.
\newblock


\bibitem[Liang et~al\mbox{.}(2021)]%
        {Liang21}
\bibfield{author}{\bibinfo{person}{Jingyun Liang}, \bibinfo{person}{Jiezhang
  Cao}, \bibinfo{person}{Guolei Sun}, \bibinfo{person}{Kai Zhang},
  \bibinfo{person}{Luc Van~Gool}, {and} \bibinfo{person}{Radu Timofte}.}
  \bibinfo{year}{2021}\natexlab{}.
\newblock \showarticletitle{SwinIR: Image Restoration Using Swin Transformer}.
  In \bibinfo{booktitle}{\emph{Proceedings of the IEEE/CVF International
  Conference on Computer Vision Workshops (ICCVW)}}.
  \bibinfo{pages}{1833--1844}.
\newblock


\bibitem[Liang et~al\mbox{.}(2022)]%
        {Liang22}
\bibfield{author}{\bibinfo{person}{Jie Liang}, \bibinfo{person}{Hui Zeng},
  {and} \bibinfo{person}{Lei Zhang}.} \bibinfo{year}{2022}\natexlab{}.
\newblock \showarticletitle{Details or Artifacts: A Locally Discriminative
  Learning Approach to Realistic Image Super-Resolution}. In
  \bibinfo{booktitle}{\emph{Proceedings of the IEEE/CVF Conference on Computer
  Vision and Pattern Recognition (CVPR)}}. \bibinfo{pages}{5657--5666}.
\newblock


\bibitem[Lim et~al\mbox{.}(2017)]%
        {Lim17}
\bibfield{author}{\bibinfo{person}{Bee Lim}, \bibinfo{person}{Sanghyun Son},
  \bibinfo{person}{Heewon Kim}, \bibinfo{person}{Seungjun Nah}, {and}
  \bibinfo{person}{Kyoung~Mu Lee}.} \bibinfo{year}{2017}\natexlab{}.
\newblock \showarticletitle{Enhanced Deep Residual Networks for Single Image
  Super-Resolution}. In \bibinfo{booktitle}{\emph{Proceedings of the IEEE/CVF
  Conference on Computer Vision and Pattern Recognition Workshops (CVPRW)}}.
  \bibinfo{pages}{1132--1140}.
\newblock


\bibitem[Liu et~al\mbox{.}(2023)]%
        {Liu23}
\bibfield{author}{\bibinfo{person}{Tao Liu}, \bibinfo{person}{Jun Cheng}, {and}
  \bibinfo{person}{Shan Tan}.} \bibinfo{year}{2023}\natexlab{}.
\newblock \showarticletitle{Spectral Bayesian Uncertainty for Image
  Super-Resolution}. In \bibinfo{booktitle}{\emph{Proceedings of the IEEE/CVF
  Conference on Computer Vision and Pattern Recognition (CVPR)}}.
  \bibinfo{pages}{18166--18175}.
\newblock


\bibitem[Liu et~al\mbox{.}(2021)]%
        {Liu21}
\bibfield{author}{\bibinfo{person}{Ze Liu}, \bibinfo{person}{Yutong Lin},
  \bibinfo{person}{Yue Cao}, \bibinfo{person}{Han Hu}, \bibinfo{person}{Yixuan
  Wei}, \bibinfo{person}{Zheng Zhang}, \bibinfo{person}{Stephen Lin}, {and}
  \bibinfo{person}{Baining Guo}.} \bibinfo{year}{2021}\natexlab{}.
\newblock \showarticletitle{Swin Transformer: Hierarchical Vision Transformer
  using Shifted Windows}. In \bibinfo{booktitle}{\emph{Proceedings of the
  IEEE/CVF International Conference on Computer Vision (ICCV)}}.
  \bibinfo{pages}{9992--10002}.
\newblock


\bibitem[Lusine et~al\mbox{.}(2022)]%
        {Lusine22}
\bibfield{author}{\bibinfo{person}{Abrahamyan Lusine}, \bibinfo{person}{Truong
  Anh~Minh}, \bibinfo{person}{Philips Wilfried}, {and}
  \bibinfo{person}{Deligiannis Nikos}.} \bibinfo{year}{2022}\natexlab{}.
\newblock \showarticletitle{Gradient Variance Loss for Structure-Enhanced Image
  Super-Resolution}.
\newblock \bibinfo{journal}{\emph{Proceedings of the International Conference
  on Acoustics, Speech, and Signal Processing (ICASSP)}}
  (\bibinfo{year}{2022}).
\newblock


\bibitem[Ma et~al\mbox{.}(2020)]%
        {Ma20}
\bibfield{author}{\bibinfo{person}{Cheng Ma}, \bibinfo{person}{Yongming Rao},
  \bibinfo{person}{Yean Cheng}, \bibinfo{person}{Ce Chen},
  \bibinfo{person}{Jiwen Lu}, {and} \bibinfo{person}{Jie Zhou}.}
  \bibinfo{year}{2020}\natexlab{}.
\newblock \showarticletitle{Structure-Preserving Super Resolution With Gradient
  Guidance}. In \bibinfo{booktitle}{\emph{Proceedings of the IEEE/CVF
  Conference on Computer Vision and Pattern Recognition (CVPR)}}.
  \bibinfo{pages}{7766--7775}.
\newblock


\bibitem[Martin et~al\mbox{.}(2001)]%
        {Martin01}
\bibfield{author}{\bibinfo{person}{David Martin}, \bibinfo{person}{Charless
  Fowlkes}, \bibinfo{person}{Doron Tal}, {and} \bibinfo{person}{Jitendra
  Malik}.} \bibinfo{year}{2001}\natexlab{}.
\newblock \showarticletitle{A Database of Human Segmented Natural Images and
  Its Application to Evaluating Segmentation Algorithms and Measuring
  Ecological Statistics}. In \bibinfo{booktitle}{\emph{Proceedings of the
  IEEE/CVF International Conference on Computer Vision (ICCV)}},
  Vol.~\bibinfo{volume}{2}. \bibinfo{pages}{416--423}.
\newblock


\bibitem[Matsui et~al\mbox{.}(2017)]%
        {Matsui17}
\bibfield{author}{\bibinfo{person}{Yusuke Matsui}, \bibinfo{person}{Kota Ito},
  \bibinfo{person}{Yuji Aramaki}, \bibinfo{person}{Azuma Fujimoto},
  \bibinfo{person}{Toru Ogawa}, \bibinfo{person}{Toshihiko Yamasaki}, {and}
  \bibinfo{person}{Kiyoharu Aizawa}.} \bibinfo{year}{2017}\natexlab{}.
\newblock \showarticletitle{Sketch-Based Manga Retrieval Using Manga109
  Dataset}.
\newblock \bibinfo{journal}{\emph{Multimedia Tools and Applications}}
  \bibinfo{volume}{76}, \bibinfo{number}{20} (\bibinfo{date}{01 Oct}
  \bibinfo{year}{2017}), \bibinfo{pages}{21811--21838}.
\newblock


\bibitem[Navarrete~Michelini et~al\mbox{.}(2019)]%
        {Navarrete19}
\bibfield{author}{\bibinfo{person}{Pablo Navarrete~Michelini},
  \bibinfo{person}{Dan Zhu}, {and} \bibinfo{person}{Hanwen Liu}.}
  \bibinfo{year}{2019}\natexlab{}.
\newblock \showarticletitle{Multi--scale Recursive and Perception--Distortion
  Controllable Image Super--Resolution}. In
  \bibinfo{booktitle}{\emph{Proceedings of the European Conference on Computer
  Vision Workshops (ECCVW)}}. \bibinfo{pages}{3--19}.
\newblock


\bibitem[Nebro et~al\mbox{.}(2009)]%
        {Nebro09}
\bibfield{author}{\bibinfo{person}{A.J. Nebro}, \bibinfo{person}{J.J. Durillo},
  \bibinfo{person}{J. Garcia-Nieto}, \bibinfo{person}{C.A. Coello~Coello},
  \bibinfo{person}{F. Luna}, {and} \bibinfo{person}{E. Alba}.}
  \bibinfo{year}{2009}\natexlab{}.
\newblock \showarticletitle{SMPSO: A New PSO-Based Mtaheuristic for
  Multi-Objective Optimization}. In \bibinfo{booktitle}{\emph{Proceedings of
  the IEEE Symposium on Computational Intelligence in Multi-Criteria
  Decision-Making (MCDM)}}. \bibinfo{pages}{66--73}.
\newblock


\bibitem[Park et~al\mbox{.}(2023b)]%
        {Park23a}
\bibfield{author}{\bibinfo{person}{JoonKyu Park}, \bibinfo{person}{Sanghyun
  Son}, {and} \bibinfo{person}{Kyoung~Mu Lee}.}
  \bibinfo{year}{2023}\natexlab{b}.
\newblock \showarticletitle{Content-Aware Local GAN for Photo-Realistic
  Super-Resolution}. In \bibinfo{booktitle}{\emph{Proceedings of the IEEE/CVF
  International Conference on Computer Vision (ICCV)}}.
  \bibinfo{pages}{10585--10594}.
\newblock


\bibitem[Park et~al\mbox{.}(2023a)]%
        {Park23}
\bibfield{author}{\bibinfo{person}{Seung~Ho Park}, \bibinfo{person}{Young~Su
  Moon}, {and} \bibinfo{person}{Nam~Ik Cho}.} \bibinfo{year}{2023}\natexlab{a}.
\newblock \showarticletitle{Perception-Oriented Single Image Super-Resolution
  using Optimal Objective Estimation}. In \bibinfo{booktitle}{\emph{Proceedings
  of the IEEE/CVF Conference on Computer Vision and Pattern Recognition
  (CVPR)}}. \bibinfo{pages}{1725--1735}.
\newblock


\bibitem[Park et~al\mbox{.}(2018)]%
        {Park18}
\bibfield{author}{\bibinfo{person}{Seong-Jin Park}, \bibinfo{person}{Hyeongseok
  Son}, \bibinfo{person}{Sunghyun Cho}, \bibinfo{person}{Ki-Sang Hong}, {and}
  \bibinfo{person}{Seungyong Lee}.} \bibinfo{year}{2018}\natexlab{}.
\newblock \showarticletitle{SRFeat: Single Image Super-Resolution with Feature
  Discrimination}. In \bibinfo{booktitle}{\emph{Proceedings of the European
  Conference on Computer Vision (ECCV)}}. \bibinfo{pages}{455--471}.
\newblock


\bibitem[Paszke et~al\mbox{.}(2019)]%
        {Paszke19}
\bibfield{author}{\bibinfo{person}{Adam Paszke}, \bibinfo{person}{Sam Gross},
  \bibinfo{person}{Francisco Massa}, \bibinfo{person}{Adam Lerer},
  \bibinfo{person}{James Bradbury}, \bibinfo{person}{Gregory Chanan},
  \bibinfo{person}{Trevor Killeen}, \bibinfo{person}{Zeming Lin},
  {et~al\mbox{.}}} \bibinfo{year}{2019}\natexlab{}.
\newblock \showarticletitle{PyTorch: An Imperative Style, High-Performance Deep
  Learning Library}. In \bibinfo{booktitle}{\emph{Proceedings of the Advances
  in Neural Information Processing Systems (NeurIPS)}}.
  \bibinfo{pages}{8024--8035}.
\newblock


\bibitem[Rad et~al\mbox{.}(2019)]%
        {Rad19}
\bibfield{author}{\bibinfo{person}{Mohammad~Saeed Rad}, \bibinfo{person}{Behzad
  Bozorgtabar}, \bibinfo{person}{Urs-Viktor Marti}, \bibinfo{person}{Max
  Basler}, \bibinfo{person}{Hazim~Kemal Ekenel}, {and}
  \bibinfo{person}{Jean-Philippe Thiran}.} \bibinfo{year}{2019}\natexlab{}.
\newblock \showarticletitle{SROBB: Targeted Perceptual Loss for Single Image
  Super-Resolution}. In \bibinfo{booktitle}{\emph{Proceedings of the IEEE/CVF
  International Conference on Computer Vision (ICCV)}}.
  \bibinfo{pages}{2710--2719}.
\newblock


\bibitem[Sajjadi et~al\mbox{.}(2017)]%
        {Sajjadi17}
\bibfield{author}{\bibinfo{person}{Mehdi S.~M. Sajjadi},
  \bibinfo{person}{Bernhard Schölkopf}, {and} \bibinfo{person}{Michael
  Hirsch}.} \bibinfo{year}{2017}\natexlab{}.
\newblock \showarticletitle{EnhanceNet: Single Image Super-Resolution Through
  Automated Texture Synthesis}. In \bibinfo{booktitle}{\emph{Proceedings of the
  IEEE/CVF International Conference on Computer Vision (ICCV)}}.
  \bibinfo{pages}{4501--4510}.
\newblock


\bibitem[Shi et~al\mbox{.}(2016)]%
        {Shi16}
\bibfield{author}{\bibinfo{person}{Wenzhe Shi}, \bibinfo{person}{Jose
  Caballero}, \bibinfo{person}{Ferenc Huszár}, \bibinfo{person}{Johannes
  Totz}, \bibinfo{person}{Andrew~P. Aitken}, \bibinfo{person}{Rob Bishop},
  \bibinfo{person}{Daniel Rueckert}, {and} \bibinfo{person}{Zehan Wang}.}
  \bibinfo{year}{2016}\natexlab{}.
\newblock \showarticletitle{Real-Time Single Image and Video Super-Resolution
  Using an Efficient Sub-Pixel Convolutional Neural Network}. In
  \bibinfo{booktitle}{\emph{Proceedings of the IEEE/CVF Conference on Computer
  Vision and Pattern Recognition (CVPR)}}. \bibinfo{pages}{1874--1883}.
\newblock


\bibitem[Simonyan and Zisserman(2015)]%
        {Simonyan15}
\bibfield{author}{\bibinfo{person}{Karen Simonyan} {and}
  \bibinfo{person}{Andrew Zisserman}.} \bibinfo{year}{2015}\natexlab{}.
\newblock \showarticletitle{Very Deep Convolutional Networks for Large-Scale
  Image Recognition}. In \bibinfo{booktitle}{\emph{Proceedings of the
  International Conference on Learning Representations (ICLR)}}.
\newblock


\bibitem[Sun et~al\mbox{.}(2022)]%
        {Sun22}
\bibfield{author}{\bibinfo{person}{Long Sun}, \bibinfo{person}{Jinshan Pan},
  {and} \bibinfo{person}{Jinhui Tang}.} \bibinfo{year}{2022}\natexlab{}.
\newblock \showarticletitle{ShuffleMixer: An Efficient ConvNet for Image
  Super-Resolution}. In \bibinfo{booktitle}{\emph{Proceedings of the Advances
  in Neural Information Processing Systems (NeurIPS)}},
  Vol.~\bibinfo{volume}{35}. \bibinfo{pages}{17314--17326}.
\newblock


\bibitem[Timofte et~al\mbox{.}(2017)]%
        {Timofte17}
\bibfield{author}{\bibinfo{person}{Radu Timofte}, \bibinfo{person}{Eirikur
  Agustsson}, \bibinfo{person}{Luc~Van Gool}, \bibinfo{person}{Ming-Hsuan
  Yang}, \bibinfo{person}{Lei Zhang}, \bibinfo{person}{Bee Lim},
  \bibinfo{person}{Sanghyun Son}, \bibinfo{person}{Heewon Kim},
  \bibinfo{person}{Seungjun Nah}, {et~al\mbox{.}}}
  \bibinfo{year}{2017}\natexlab{}.
\newblock \showarticletitle{NTIRE 2017 Challenge on Single Image
  Super-Resolution: Methods and Results}. In
  \bibinfo{booktitle}{\emph{Proceedings of the IEEE/CVF Conference on Computer
  Vision and Pattern Recognition Workshops (CVPRW)}}.
  \bibinfo{pages}{1110--1121}.
\newblock


\bibitem[Umer and Micheloni(2020)]%
        {Umer20}
\bibfield{author}{\bibinfo{person}{Rao~Muhammad Umer} {and}
  \bibinfo{person}{Christian Micheloni}.} \bibinfo{year}{2020}\natexlab{}.
\newblock \showarticletitle{Deep Cyclic Generative Adversarial Residual
  Convolutional Networks for Real Image Super-Resolution}. In
  \bibinfo{booktitle}{\emph{Proceedings of the European Conference on Computer
  Vision (ECCV)}}. \bibinfo{pages}{484--498}.
\newblock


\bibitem[Vu et~al\mbox{.}(2019)]%
        {Vu18}
\bibfield{author}{\bibinfo{person}{Thang Vu}, \bibinfo{person}{Tung~M. Luu},
  {and} \bibinfo{person}{Chang~D. Yoo}.} \bibinfo{year}{2019}\natexlab{}.
\newblock \showarticletitle{Perception-Enhanced Image Super-Resolution via
  Relativistic Generative Adversarial Networks}. In
  \bibinfo{booktitle}{\emph{Proceedings of the European Conference on Computer
  Vision Workshops (ECCVW)}}. \bibinfo{pages}{98--113}.
\newblock


\bibitem[Wagner et~al\mbox{.}(2010)]%
        {Wagner10}
\bibfield{author}{\bibinfo{person}{Tobias Wagner}, \bibinfo{person}{Michael
  Emmerich}, \bibinfo{person}{Andr{\'e} Deutz}, {and} \bibinfo{person}{Wolfgang
  Ponweiser}.} \bibinfo{year}{2010}\natexlab{}.
\newblock \showarticletitle{On Expected-Improvement Criteria for Model-based
  Multi-objective Optimization}. In \bibinfo{booktitle}{\emph{Proceedings of
  the Parallel Problem Solving from Nature (PPSN)}}. \bibinfo{pages}{718--727}.
\newblock


\bibitem[Wang et~al\mbox{.}(2023)]%
        {Wang23}
\bibfield{author}{\bibinfo{person}{Chenyang Wang}, \bibinfo{person}{Junjun
  Jiang}, \bibinfo{person}{Zhiwei Zhong}, {and} \bibinfo{person}{Xianming
  Liu}.} \bibinfo{year}{2023}\natexlab{}.
\newblock \showarticletitle{Spatial-Frequency Mutual Learning for Face
  Super-Resolution}. In \bibinfo{booktitle}{\emph{Proceedings of the IEEE/CVF
  Conference on Computer Vision and Pattern Recognition (CVPR)}}.
  \bibinfo{pages}{22356--22366}.
\newblock


\bibitem[Wang et~al\mbox{.}(2019a)]%
        {Wang19a}
\bibfield{author}{\bibinfo{person}{Wei Wang}, \bibinfo{person}{Ruiming Guo},
  \bibinfo{person}{Yapeng Tian}, {and} \bibinfo{person}{Wenming Yang}.}
  \bibinfo{year}{2019}\natexlab{a}.
\newblock \showarticletitle{CFSNet: Toward a Controllable Feature Space for
  Image Restoration}. In \bibinfo{booktitle}{\emph{Proceedings of the IEEE/CVF
  International Conference on Computer Vision (ICCV)}}.
  \bibinfo{pages}{4139--4148}.
\newblock


\bibitem[Wang et~al\mbox{.}(2018)]%
        {Wang18a}
\bibfield{author}{\bibinfo{person}{Xintao Wang}, \bibinfo{person}{Ke Yu},
  \bibinfo{person}{Chao Dong}, {and} \bibinfo{person}{Chen Change~Loy}.}
  \bibinfo{year}{2018}\natexlab{}.
\newblock \showarticletitle{Recovering Realistic Texture in Image
  Super-Resolution by Deep Spatial Feature Transform}. In
  \bibinfo{booktitle}{\emph{Proceedings of the IEEE/CVF Conference on Computer
  Vision and Pattern Recognition (CVPR)}}. \bibinfo{pages}{606--615}.
\newblock


\bibitem[Wang et~al\mbox{.}(2019b)]%
        {Wang19b}
\bibfield{author}{\bibinfo{person}{Xintao Wang}, \bibinfo{person}{Ke Yu},
  \bibinfo{person}{Chao Dong}, \bibinfo{person}{Xiaoou Tang}, {and}
  \bibinfo{person}{Chen~Change Loy}.} \bibinfo{year}{2019}\natexlab{b}.
\newblock \showarticletitle{Deep Network Interpolation for Continuous Imagery
  Effect Transition}. In \bibinfo{booktitle}{\emph{Proceedings of the IEEE/CVF
  Conference on Computer Vision and Pattern Recognition (CVPR)}}.
  \bibinfo{pages}{1692--1701}.
\newblock


\bibitem[Wang et~al\mbox{.}(2019c)]%
        {Wang19}
\bibfield{author}{\bibinfo{person}{Xintao Wang}, \bibinfo{person}{Ke Yu},
  \bibinfo{person}{Shixiang Wu}, \bibinfo{person}{Jinjin Gu},
  \bibinfo{person}{Yihao Liu}, \bibinfo{person}{Chao Dong}, \bibinfo{person}{Yu
  Qiao}, {and} \bibinfo{person}{Chen~Change Loy}.}
  \bibinfo{year}{2019}\natexlab{c}.
\newblock \showarticletitle{ESRGAN: Enhanced Super-Resolution Generative
  Adversarial Networks}. In \bibinfo{booktitle}{\emph{Proceedings of the
  European Conference on Computer Vision (ECCV)}}. \bibinfo{pages}{63--79}.
\newblock


\bibitem[Wang(2022)]%
        {Wang22}
\bibfield{author}{\bibinfo{person}{Yan Wang}.} \bibinfo{year}{2022}\natexlab{}.
\newblock \showarticletitle{Edge-enhanced Feature Distillation Network for
  Efficient Super-Resolution}. In \bibinfo{booktitle}{\emph{Proceedings of the
  IEEE/CVF Conference on Computer Vision and Pattern Recognition Workshops
  (CVPRW)}}. \bibinfo{pages}{776--784}.
\newblock


\bibitem[Wang et~al\mbox{.}(2004)]%
        {Zhou04}
\bibfield{author}{\bibinfo{person}{Zhou Wang}, \bibinfo{person}{A.C. Bovik},
  \bibinfo{person}{H.R. Sheikh}, {and} \bibinfo{person}{E.P. Simoncelli}.}
  \bibinfo{year}{2004}\natexlab{}.
\newblock \showarticletitle{Image Quality Assessment: from Error Visibility to
  Structural Similarity}.
\newblock \bibinfo{journal}{\emph{IEEE Transactions on Image Processing}}
  \bibinfo{volume}{13}, \bibinfo{number}{4} (\bibinfo{year}{2004}),
  \bibinfo{pages}{600--612}.
\newblock


\bibitem[Zeyde et~al\mbox{.}(2012)]%
        {Zeyde12}
\bibfield{author}{\bibinfo{person}{Roman Zeyde}, \bibinfo{person}{Michael
  Elad}, {and} \bibinfo{person}{Matan Protter}.}
  \bibinfo{year}{2012}\natexlab{}.
\newblock \showarticletitle{On Single Image Scale-Up Using
  Sparse-Representations}. In \bibinfo{booktitle}{\emph{Proceedings of the
  International Conference on Curves and Surfaces}}. \bibinfo{pages}{711--730}.
\newblock


\bibitem[Zhang and Li(2007)]%
        {Zhang07}
\bibfield{author}{\bibinfo{person}{Qingfu Zhang} {and} \bibinfo{person}{Hui
  Li}.} \bibinfo{year}{2007}\natexlab{}.
\newblock \showarticletitle{MOEA/D: A Multiobjective Evolutionary Algorithm
  Based on Decomposition}.
\newblock \bibinfo{journal}{\emph{IEEE Transactions on Evolutionary
  Computation}} \bibinfo{volume}{11}, \bibinfo{number}{6}
  (\bibinfo{year}{2007}), \bibinfo{pages}{712--731}.
\newblock


\bibitem[Zhang et~al\mbox{.}(2018)]%
        {Zhang18}
\bibfield{author}{\bibinfo{person}{Richard Zhang}, \bibinfo{person}{Phillip
  Isola}, \bibinfo{person}{Alexei~A. Efros}, \bibinfo{person}{Eli Shechtman},
  {and} \bibinfo{person}{Oliver Wang}.} \bibinfo{year}{2018}\natexlab{}.
\newblock \showarticletitle{The Unreasonable Effectiveness of Deep Features as
  a Perceptual Metric}. In \bibinfo{booktitle}{\emph{Proceedings of the
  IEEE/CVF Conference on Computer Vision and Pattern Recognition (CVPR)}}.
  \bibinfo{pages}{586--595}.
\newblock


\bibitem[Zitzler et~al\mbox{.}(2001)]%
        {Zitzler01}
\bibfield{author}{\bibinfo{person}{Eckart Zitzler}, \bibinfo{person}{Marco
  Laumanns}, {and} \bibinfo{person}{Lothar Thiele}.}
  \bibinfo{year}{2001}\natexlab{}.
\newblock \showarticletitle{SPEA2: Improving the strength pareto evolutionary
  algorithm}.
\newblock \bibinfo{journal}{\emph{Technical Report Gloriastrasse}}
  (\bibinfo{year}{2001}).
\newblock


\end{thebibliography}
	
\end{document}

\title{Supplementary Materials: Perceptual Oriented Image Restoration is a Multi-Objective Optimization Problem}

	\author{Qiwen Zhu}
	\orcid{0009-0008-5984-4941}
	\affiliation{
		\institution{State Key Lab of MIIPT, Huazhong University of Science and Technology}
		\city{Wuhan}
		\country{China}
	}
	\email{zhuqiwen@hust.edu.cn}
	\author{Yanjie Wang}
	\orcid{0000-0001-7352-6183}
	\affiliation{
		\institution{School of AIA, Huazhong University of Science and Technology}
		\city{Wuhan}
		\country{China}
	}
	\email{aiawyj@hust.edu.cn}
	\author{Shilv Cai}
	\orcid{0000-0002-4037-4555}
	\affiliation{
		\institution{School of AIA, Huazhong University of Science and Technology}
		\city{Wuhan}
		\country{China}
	}
	\email{caishilv@hust.edu.cn}
	\author{Liqun Chen}
	\orcid{0000-0001-6345-5713}
	\affiliation{
		\institution{School of AIA, Huazhong University of Science and Technology}
		\city{Wuhan}
		\country{China}
	}
	\email{chenliqun@hust.edu.cn}
	\author{Jiahuan Zhou}
	\orcid{0000-0002-3301-747X}
	\affiliation{
		\institution{Wangxuan Institute of Computer Technology, Peking University}
		\city{Beijing}
		\country{China}
	}
	\email{jiahuanzhou@pku.edu.cn}
	\author{Luxin Yan}
	\orcid{0000-0002-5445-2702}
	\affiliation{
		\institution{State Key Lab of MIIPT, Huazhong University of Science and Technology}
		\city{Wuhan}
		\country{China}
	}
	\email{yanluxin@hust.edu.cn}
	\author{Sheng Zhong}
	\orcid{0000-0003-2865-8202}
	\affiliation{
		\institution{State Key Lab of MIIPT, Huazhong University of Science and Technology}
		\city{Wuhan}
		\country{China}
	}
	\email{zhongsheng@hust.edu.cn}
	\author{Xu Zou}
	\orcid{0000-0002-0251-7404}
	\authornote{Corresponding author.}
	\affiliation{
		\institution{State Key Lab of MIIPT, Huazhong University of Science and Technology}
		\city{Wuhan}
		\country{China}
	}
	\email{zoux@hust.edu.cn}
	
	\maketitle
	
	\begin{figure}[htbp]
		\centering
		\includegraphics[width=\linewidth]{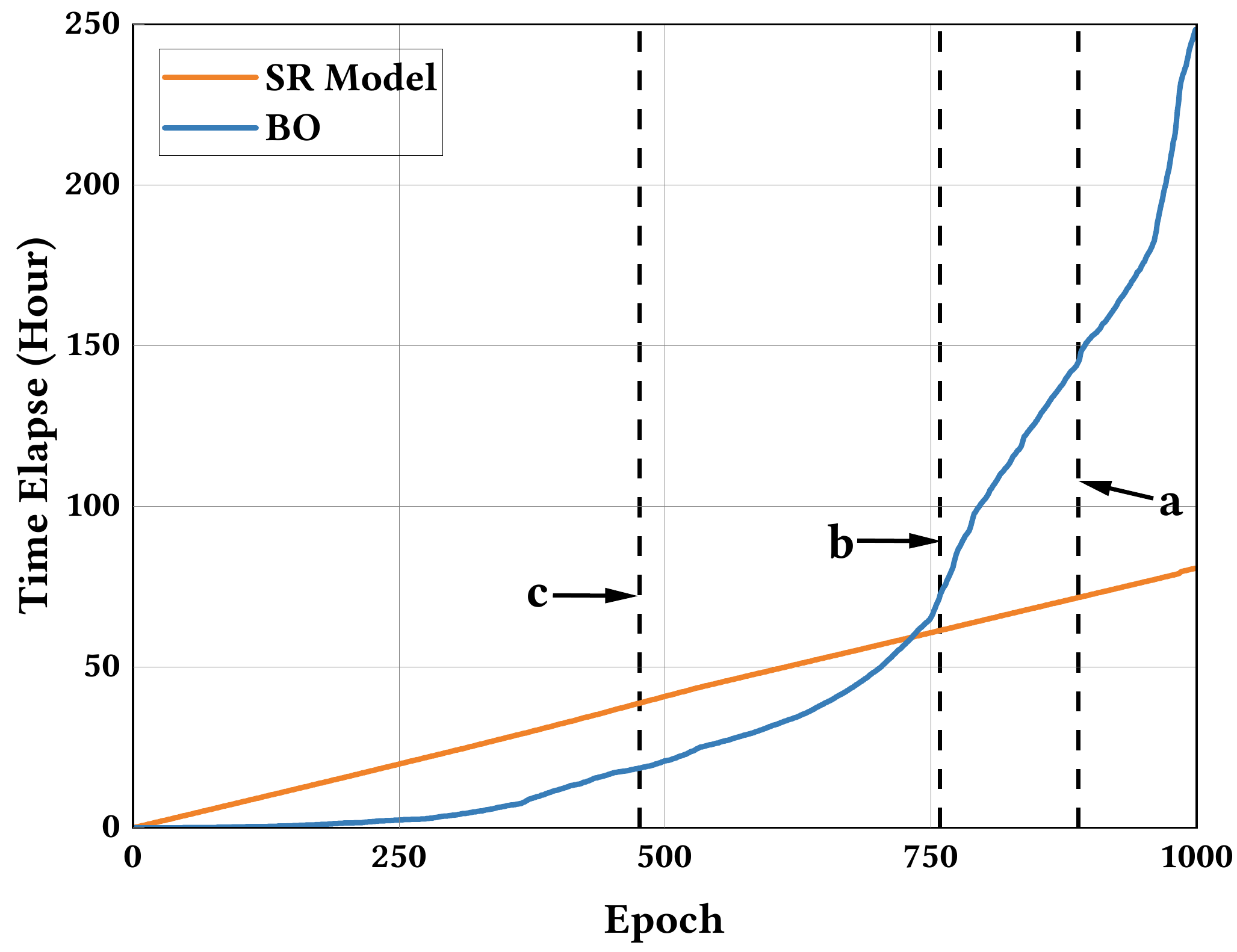}
		\caption{Comparative of the time consumed for training the SR model against the time required for BO during the training process of our MOBOSR. The temporal measurements were conducted on an server with NVIDIA RTX 3090 GPU and two Intel Xeon Gold 6226R CPUs. Epochs for Ours-[a,b,c] are labeled for clarity.}
		\label{fig:time_elapse}
	\end{figure}
	
	\begin{table}[htbp]
		\caption{Detailed comparison of time expended for sampling points Ours-[a,b,c] and at the optimization cessation.}
		\label{tab:time_elapse}
		\begin{tabular}{cccc}
			\toprule
			Point & Epoch & SR Time Elapse & MOBO Time Elapse \\
			\midrule
			Ours-c & 476 & 38.8h & 18.6h \\
			Ours-b & 759 & 61.4h & 72.6h \\
			Ours-a & 889 & 71.7h & 145.2h \\
			End & 1000 & 80.8h & 248.4h \\
			\bottomrule
		\end{tabular}
	\end{table}
	
	\section{Time Analysis}
	
	Bayesian Optimization (BO) minimizes the number of evaluations by substituting the actual objective function with a surrogate function and heuristically determining the most promising points for improvement through an acquisition function for subsequent evaluation rounds. Consequently, this approach significantly conserves optimization iteration compared to evolutionary algorithms. However, the BO process necessitates the computation of the covariance matrix, resulting in a time complexity of $O(n^3)$. If the objective function be overly intricate or the problem dimensions too high, necessitating numerous optimization rounds, this would substantially increase the optimization time.
	
	\begin{figure}[htbp]
		\centering
		\includegraphics[width=\linewidth]{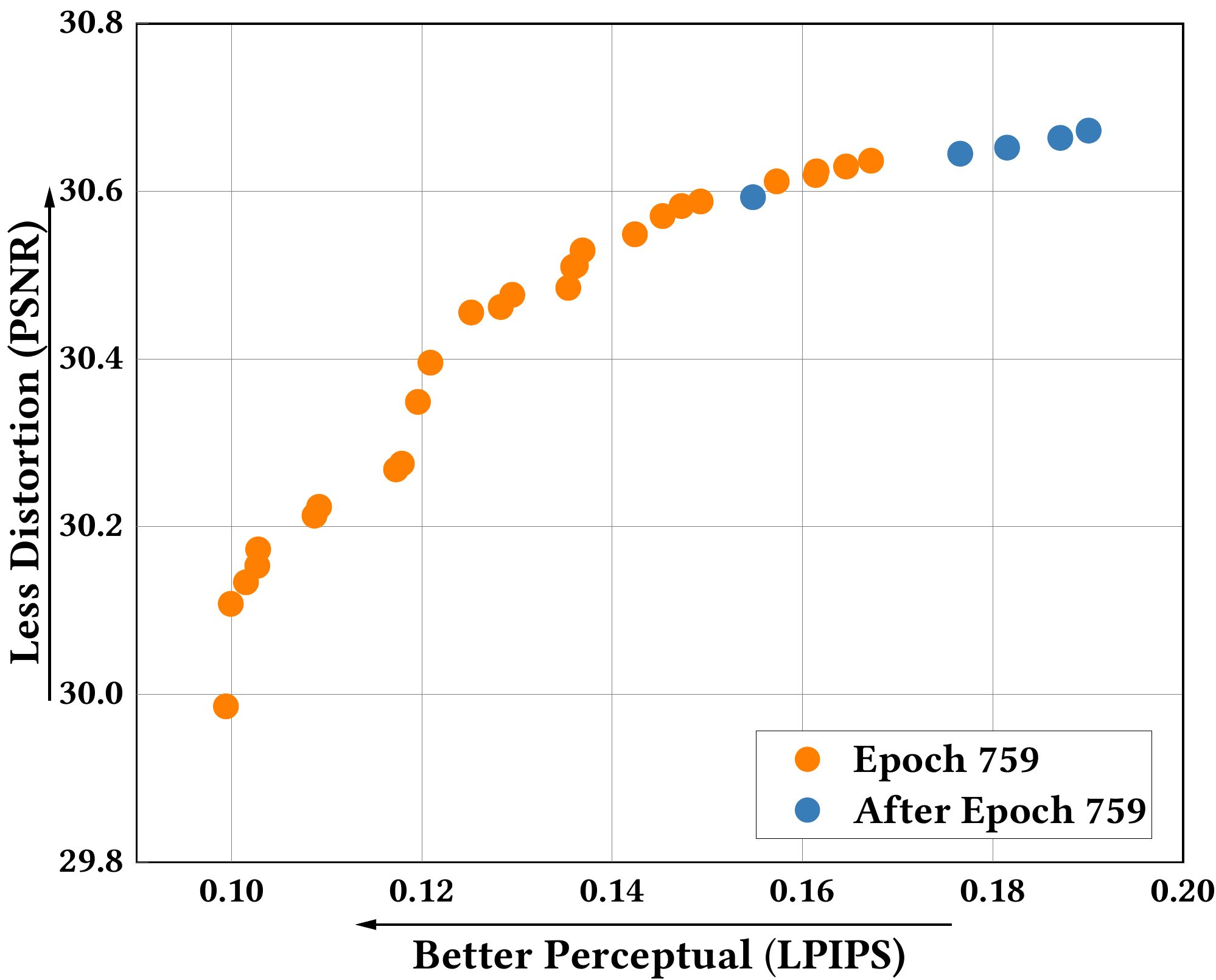}
		\caption{Changes in the perception-distortion Pareto frontier obtained by our MOBOSR after epoch 759.}
		\label{fig:pareto_front_759}
	\end{figure}
	
	In the main text, the three sampling points selected from the perceptual-distortion Pareto front, labeled as Ours-[a,b,c], were obtained at epochs 476, 759 and 889, respectively. As shown in Table \ref{tab:time_elapse} and Figure \ref{fig:time_elapse}, the time consumed to train the Super-Resolution (SR) model with Ours-c, obtained at epoch 476, was approximately half of that required for BO. At epoch 759, the training time for the SR model with Ours-b was nearly identical to that for BO. However, by epoch 889, the time expended on BO for Ours-a was roughly half of that for training the SR model. By the designated optimization halt at epoch 1000, the time consumption for BO had become threefold that of the SR model.
	
	But, after optimization up to epoch 759 (Ours-c), there were no significant changes in the Pareto front (as shown in Figure \ref{fig:pareto_front_759}). We believe that the substantial advantages brought by our Multi-Objective Bayesian Optimization Super-Resolution (MOBOSR), and without introducing any additional computational load during inference, justify the mere doubling of training duration.
	
	\section{Discussion on Multi-Task Learning}
	
	\begin{table}[t]
		\caption{Analyses of multi-task learning on the DIV2K \cite{Agustsson17} validation set. The best and second-best results are highlighted in \textbf{bold} and \underline{underline}, respectively. MTL methods show significant improvements over the manually weighted ESRGAN \cite{Wang19} but are not as effective as our MOBOSR.}
		\label{tab:multi_task}
		\begin{tabular*}{\linewidth}{@{\extracolsep{\fill}}c|cccc}
			\toprule
			Methods & PSNR$\uparrow$ & SSIM$\uparrow$ & LR-PSNR$\uparrow$ & LPIPS$\downarrow$ \\
			\midrule
			ESRGAN \cite{Wang19} & 27.6994 & 0.7610 & 41.2244 & 0.1193 \\
			CAGrad \cite{Liu21a} & \underline{28.4434} & \underline{0.7803} & \textbf{47.5291} & \underline{0.1161} \\
			NashMTL \cite{Navon22} & 28.1606 & 0.7717 & 43.7037 & 0.1168 \\
			MOBOSR (Ours) & \textbf{28.5089} & \textbf{0.7834} & \underline{44.6847} & \textbf{0.1145} \\
			\bottomrule
		\end{tabular*}
	\end{table}
	
	Focusing on a single task may overlook the information from related tasks that could improve the target task. By sharing parameters between different tasks~(with different loss functions) to a certain extent may achieve better generalization for the original task. Hence, we discuss the impact of Multi-Task Learning (MTL) in balancing the distortion and perceptual quality for SISR models. We initially consider using MTL to optimize multiple loss functions to address the balance issue. We compare the results of our MOBOSR with 2 MTL methods~(CAGrad \cite{Liu21a} and NashMTL \cite{Navon22}), as well as with ESRGAN \cite{Wang19}. As shown in Table \ref{tab:multi_task}, MTL methods show significant improvements over the manually weighted ESRGAN \cite{Wang19} but are not as effective as our MOBOSR. We believe this is because MTL approaches just seek a single compromise between multi-task/multi-loss rather than searching for the entire Pareto frontier as MOBO does. But, the ability to achieve these outcomes indicates that MTL merits further investigation.
	
	\section{Metrics Recalculation Details}
	
	\begin{table*}
		\caption{Comparison of metrics calculated using our uniform method versus those reported by the authors, showing minimal difference, with some even surpassing the reported results. The higher results are highlighted in \textbf{bold}. The symbols $\uparrow$ and $\downarrow$ indicate that higher or lower values of the metric are preferable.}
		\label{tab:metric_compare}
		\resizebox{\textwidth}{!}{
			\begin{tabular}{cl|cc|cc|cc|cc|cc}
				\toprule
				\multirow{2}*{Metric} & \multirow{2}*{Dataset} & \multicolumn{2}{c|}{RRDB-PSNR \cite{Wang19}} & \multicolumn{2}{c|}{SPSR \cite{Ma20}} & \multicolumn{2}{c|}{RRDB+LDL \cite{Liang22}} & \multicolumn{2}{c}{CAL-GAN \cite{Park23a}} & \multicolumn{2}{c}{SROOE \cite{Park23}} \\
				& & Author & Recalculated & Author & Recalculated & Author & Recalculated & Author & Recalculated & Author & Recalculated \\
				\midrule
				\multirow{7}*{PSNR$\uparrow$}
				& Set5 & \textbf{32.73} & 32.7010 & \textbf{30.400} & 30.3871 & 30.985 & \textbf{31.0007} & \textbf{31.177} & 31.0475 & - & 31.2455 \\
				& Set14 & \textbf{28.99} & 28.9831 & 26.640 & \textbf{26.6501} & \textbf{27.491} & 27.2064 & - & 27.3272 & - & 27.2561 \\
				& DIV2K & - & 30.8888 & - & 28.1824 & 28.951 & \textbf{28.9510} & 28.863 & \textbf{28.9549} & 27.69 & \textbf{29.0990} \\
				& BSD100 & \textbf{27.85} & 27.8235 & \textbf{25.505} & 25.4949 & - & 26.0988 & 25.925 & \textbf{26.2581} & 24.87 & \textbf{26.1715} \\
				& Urban100 & \textbf{27.03} & 26.9859 & 24.799 & \textbf{24.8063} & \textbf{25.498} & 25.4781 & 25.290 & \textbf{25.2908} & 24.33 & \textbf{25.8452}\\
				& General100 & - & 31.9145 & 29.414 & \textbf{29.4794} & \textbf{30.232} & 30.1974 & \textbf{30.182} & 30.0742 & 28.74 & \textbf{30.4723}\\
				& Manga109 & \textbf{31.66} & 31.5637 & - & 28.6102 & 29.407 & \textbf{29.4111} & - & 29.1665 & 28.08 & \textbf{29.9017} \\
				\hline
				\multirow{7}*{SSIM$\uparrow$}
				& Set5 & \textbf{0.9011} & 0.9010 & \textbf{0.8627} & 0.8432  & \textbf{0.8626} & 0.8610  & \textbf{0.863} & 0.8552 & - & 0.8651 \\
				& Set14 & \textbf{0.7917} & 0.7915 & \textbf{0.7930} & 0.7133 & \textbf{0.7476} & 0.7343 & - & 0.7353 & - & 0.7304 \\
				& DIV2K & - & 0.8485 & - & 0.7720  & 0.7951 & \textbf{0.7952} & \textbf{0.790} & 0.7897 & 0.7932 & \textbf{0.7980} \\
				& BSD100 & \textbf{0.7455} & 0.7453 & \textbf{0.6576} & 0.6571 & - & 0.6811 & 0.676 & \textbf{0.6789} & \textbf{0.6869} & 0.6866 \\
				& Urban100 & \textbf{0.8153} & 0.8152 & \textbf{0.9481} & 0.7472 & \textbf{0.7673} & 0.7670 & \textbf{0.763} & 0.7623 & 0.7707 & \textbf{0.7764} \\
				& General100 & - & 0.8725 & \textbf{0.8537} & 0.8095 & 0.8277  & \textbf{0.8278} & 0.825 & \textbf{0.8262} & 0.8297 & \textbf{0.8332} \\
				& Manga109 & \textbf{0.9196} & 0.9195 & - & 0.8591 & 0.8746 & \textbf{0.8746} & - & 0.8676 & 0.8554 & \textbf{0.8786} \\
				\hline
				\multirow{7}*{LR-PSNR$\uparrow$}
				& Set5 & - & 53.0951 & - & 46.3607 & - & 48.5067 & - & 42.4327 & - & 53.1781 \\
				& Set14 & - & 50.8098 & - & 43.6201 & - & 46.2893 & - & 41.5963 & - & 51.0679 \\
				& DIV2K & - & 51.9030 & - & 44.8529 & - & 47.9757 & - & 42.8611 & 50.80 & \textbf{53.5488} \\
				& BSD100 & - & 50.3975 & - & 42.6756 & - & 45.1571 & - & 41.0666 & 49.19 & \textbf{51.2347} \\
				& Urban100 & - & 51.0009 & - & 42.6679 & - & 46.5827 & - & 41.6069 & 48.32 & \textbf{50.6700} \\
				& General100 & - & 52.6741 & - & 44.6786 & - & 48.0079 & - & 43.4227 & 50.11 & \textbf{52.9797} \\
				& Manga109 & - & 52.3321 & - & 44.3872 & - & 47.8923 & - & 42.8636 & 48.77 & \textbf{51.7820} \\
				\hline
				\multirow{7}*{LPIPS$\downarrow$}
				& Set5 & - & 0.1691 & 0.0644 & \textbf{0.0616} & 0.0670 & \textbf{0.0637} & \textbf{0.061} & 0.0687 & - & 0.0603 \\
				& Set14 & - & 0.2718 & 0.1318 & \textbf{0.1313} & \textbf{0.1207} & 0.1309 & - & 0.1320 & - & 0.1131 \\
				& DIV2K & - & 0.2526 & - & 0.1097 & 0.1011 & \textbf{0.1007} & \textbf{0.091} & 0.1072 & 0.0957 & \textbf{0.0956} \\
				& BSD100 & - & 0.3590 & \textbf{0.1611} & 0.1629 & - & 0.1635 & \textbf{0.151} & 0.1696 & \textbf{0.1500} & 0.1514 \\
				& Urban100 & - & 0.1956 & \textbf{0.1184} & 0.1186 & \textbf{0.1096} & 0.1097 & \textbf{0.108} & 0.1171 & \textbf{0.1065} & 0.1067 \\
				& General100 & - & 0.1668 & \textbf{0.0863} & 0.0866 & \textbf{0.0790} & 0.0794 & \textbf{0.077} & 0.0894 & \textbf{0.0753} & 0.0758 \\
				& Manga109 & - & 0.0977 & - & 0.0662 & \textbf{0.0553} & 0.0546 & - & 0.0688 & 0.0524 & \textbf{0.0511} \\
				\bottomrule
		\end{tabular}}
	\end{table*}
	
	Due to the variations in datasets and metrics reported by the methods under comparison, as well as the differences in implementation details during metric computation, we have adopted a uniform metric calculation method to re-evaluate the metrics of other methods. This ensures a fairer comparison. We generated SR results for all test sets using the model weights and inference code released by the authors, followed by calculations using our standardized metric computation program. Our codes for metric calculation are detailed in the GitHub repository: \href{https://github.com/ZhuKeven/MOBOSR}{https://github.com/ZhuKeven/MOBOSR}. Table \ref{tab:metric_compare} presents the metrics we recomputed alongside those reported by the authors, with most showing no significant differences and some even surpassing the reported results. Wang et al. only reported metrics for the RRDB-PSNR \cite{Wang19} model trained with L1 loss, without providing the metrics for the ESRGAN \cite{Wang19} model trained using GAN \cite{Goodfellow20}. Consequently, we are unable to present the ESRGAN \cite{Wang19} metrics comparison in Table \ref{tab:metric_compare} in the same manner as for other methods.
	
	\section{More Quantitative Results}
	
	\begin{table*}
		\caption{Comparison of Ours-[a,b,c] with other artworks on 7 datasets. The best, second-best and third-best results are highlighted in \textbf{bold}, \underline{underline} and \textit{italic}, respectively. The symbols $\uparrow$ and $\downarrow$ indicate that higher or lower values of the metric are preferable.}
		\label{tab:compare_2_sota}
		\begin{tabular*}{\textwidth}{@{\extracolsep{\fill}}c|c|c|ccccccc}
			\toprule
			Metric & Method & Train Datasets & Set5 & Set14 & DIV2K & BSD100 & Urban100 & General100 & Manga109 \\
			\midrule
			\multirow{9}*{PSNR$\uparrow$}
			& RRDB-PSNR \cite{Wang19} & DF2K-OST & \textbf{32.7010} & \textbf{28.9831} & \textbf{30.8888} & \textbf{27.8235} & \textbf{26.9859} & \textbf{31.9145 }& \textbf{31.5637} \\
			& ESRGAN \cite{Wang19} & DF2K-OST & 30.4618 & 26.2839 & 28.1778 & 25.2892 & 24.3617 & 29.4593 & 28.5041 \\
			& SPSR \cite{Ma20} & DIV2K & 30.3871 & 26.6501 & 28.1824 & 25.4949 & 24.8063 & 29.4794 & 28.6102 \\
			& RRDB+LDL \cite{Liang22} & DIV2K & 31.0007 & 27.2064 & 28.9510 & 26.0988 & 25.4781 & 30.1974 & 29.4111 \\
			& CAL-GAN \cite{Park23a} & DIV2K & 31.0475 & 27.3272 & 28.9549 & 26.2581 & 25.2908 & 30.0742 & 29.1665 \\
			& SROOE \cite{Park23} & DF2K & 31.2455 & 27.2561 & 29.0990 & 26.1715 & 25.8452 & 30.4723 & 29.9017 \\
			& Ours-a & DIV2K & \underline{32.3663} & \underline{28.7621} & \underline{30.6384} & \underline{27.6546} & \underline{26.5285} & \underline{31.6047} & \underline{30.9787} \\
			& Ours-b & DIV2K & \textit{32.2126} & \textit{28.6426} & \textit{30.4890} & \textit{27.5176} & \textit{26.4439} & \textit{31.4769} & \textit{30.8840} \\
			& Ours-c & DIV2K & 31.8272 & 28.1766 & 29.9858 & 27.0494 & 26.0764 & 31.1164 & 30.2763 \\
			\hline
			\multirow{9}*{SSIM$\uparrow$}
			& RRDB-PSNR \cite{Wang19} & DF2K-OST & \textbf{0.9010} & \textbf{0.7915} & \textbf{0.8485} & \textbf{0.7453} & \textbf{0.8152} & \textbf{0.8725} & \textbf{0.9195} \\
			& ESRGAN \cite{Wang19} & DF2K-OST & 0.8518  & 0.6982  & 0.7761  & 0.6496  & 0.7341  & 0.8102  & 0.8604  \\
			& SPSR \cite{Ma20} & DIV2K & 0.8432  & 0.7133  & 0.7720  & 0.6571  & 0.7472  & 0.8095  & 0.8591  \\
			& RRDB+LDL \cite{Liang22} & DIV2K & 0.8610 & 0.7343  & 0.7952  & 0.6811  & 0.7670  & 0.8278 & 0.8746  \\
			& CAL-GAN \cite{Park23a} & DIV2K & 0.8552 & 0.7353  & 0.7897  & 0.6789  & 0.7623  & 0.8262  & 0.8676  \\
			& SROOE \cite{Park23} & DF2K & 0.8651 & 0.7304 & 0.7980 & 0.6866 & 0.7764 & 0.8332 & 0.8786 \\
			& Ours-a & DIV2K & \underline{0.8961} & \underline{0.7847} & \underline{0.8417} & \underline{0.7379} & \underline{0.7989} & \underline{0.8665} & \underline{0.9133} \\
			& Ours-b & DIV2K & \textit{0.8918} & \textit{0.7800} & \textit{0.8376} & \textit{0.7323} & \textit{0.7963} & \textit{0.8629} & \textit{0.9093} \\
			& Ours-c & DIV2K & 0.8804 & 0.7615 & 0.8203 & 0.7109 & 0.7812 & 0.8495 & 0.8938 \\
			\hline
			\multirow{9}*{LR-PSNR$\uparrow$}
			& RRDB-PSNR \cite{Wang19} & DF2K-OST & 53.0951 & 50.8098 & 51.9030 & 50.3975 & 51.0009 & 52.6741 & 52.3321 \\
			& ESRGAN \cite{Wang19} & DF2K-OST & 46.7348 & 43.8433 & 45.9012 & 43.8190 & 42.9339 & 45.4220 & 43.9667 \\
			& SPSR \cite{Ma20} & DIV2K & 46.3607 & 43.6201 & 44.8529 & 42.6756 & 42.6679 & 44.6786 & 44.3872 \\
			& RRDB+LDL \cite{Liang22} & DIV2K & 48.5067 & 46.2893 & 47.9757 & 45.1571 & 46.5827 & 48.0079 & 47.8923 \\
			& CAL-GAN \cite{Park23a} & DIV2K & 42.4327 & 41.5963 & 42.8611 & 41.0666 & 41.6069 & 43.4227 & 42.8636 \\
			& SROOE \cite{Park23} & DF2K & 53.1781 & 51.0679 & 53.5488 & 51.2347 & 50.6700 & 52.9797 & 51.7820 \\
			& Ours-a & DIV2K & \underline{53.7806} & \textbf{53.5768} & \underline{54.7418} & \textbf{54.2712} & \textbf{53.7559} & \underline{54.1618} & \underline{53.8049} \\
			& Ours-b & DIV2K & \textit{53.4204} & \textit{53.1927} & \textit{53.9850} & \textit{53.3449} & \underline{53.3426} & \textit{53.9097} & \textbf{53.9045} \\
			& Ours-c & DIV2K & \textbf{54.3372} & \underline{53.3344} & \textbf{55.2161} & \underline{53.3618} & \textit{52.9401} & \textbf{54.5283} & \textit{53.4195} \\
			\hline
			\multirow{9}*{LPIPS$\downarrow$}
			& RRDB-PSNR \cite{Wang19} & DF2K-OST & 0.1691 & 0.2718 & 0.2526 & 0.3590 & 0.1956 & 0.1668 & 0.0977 \\
			& ESRGAN \cite{Wang19} & DF2K-OST & 0.0750  & 0.1341  & 0.1155  & \textit{0.1617} & 0.1228  & 0.0876  & 0.0647  \\
			& SPSR \cite{Ma20} & DIV2K & \textit{0.0616}  & 0.1313  & 0.1097  & 0.1629  & 0.1186  & 0.0866  & 0.0662  \\
			& RRDB+LDL \cite{Liang22} & DIV2K & 0.0637  & \textit{0.1309} & \textit{0.1007}  & 0.1635  & \underline{0.1097}  & \textit{0.0794}  & \underline{0.0546}  \\
			& CAL-GAN \cite{Park23a} & DIV2K & 0.0687  & 0.1320  & 0.1072  & 0.1696  & 0.1171  & 0.0894  & 0.0688  \\
			& SROOE \cite{Park23} & DF2K & \textbf{0.0603} & \textbf{0.1131} & \textbf{0.0956} & \underline{0.1514} & \textbf{0.1067} & \textbf{0.0758} & \textbf{0.0511} \\
			& Ours-a & DIV2K & 0.1293 & 0.2148 & 0.1887 & 0.2723 & 0.1811 & 0.1342 & 0.0794 \\
			& Ours-b & DIV2K & 0.0849 & 0.1604 & 0.1365 & 0.2051 & 0.1432 & 0.0978 & 0.0598 \\
			& Ours-c & DIV2K & \underline{0.0607}  & \underline{0.1240}  & \underline{0.0994}  & \textbf{0.1508}  & \textit{0.1154}  & \underline{0.0776}  & \textit{0.0576}  \\
			\bottomrule
		\end{tabular*}
	\end{table*}
	
	Due to the length constraints of the main text, we have included the complete results for Ours-[a,b,c] here, as well as the metrics for the RRDB-PSNR \cite{Wang19} model trained using L1 loss. Although RRDB-PSNR \cite{Wang19} exhibits superior performance in terms of PSNR and SSIM \cite{Zhou04} metrics, the margin by which it surpasses Ours-a is considerably less than the extent to which Ours-a exceeds RRDB-PSNR \cite{Wang19} in perceptual (LPIPS) and consistency (LR-PSNR) metrics. Not to mention that the RRDB-PSNR \cite{Wang19} was trained on a significantly larger dataset, the DF2K-OST (13774 images), which comprises the DIV2K \cite{Agustsson17} training set (800 images), Flickr2K \cite{Timofte17} (2650 images), and OST \cite{Wang18a} (10,324 images), whereas our MOBOSR was trained solely on the DIV2K \cite{Agustsson17} training set (800 images).
	
	\section{More Visual Results}
	
	\begin{figure*}[h]
		\centering
		\includegraphics[width=\textwidth]{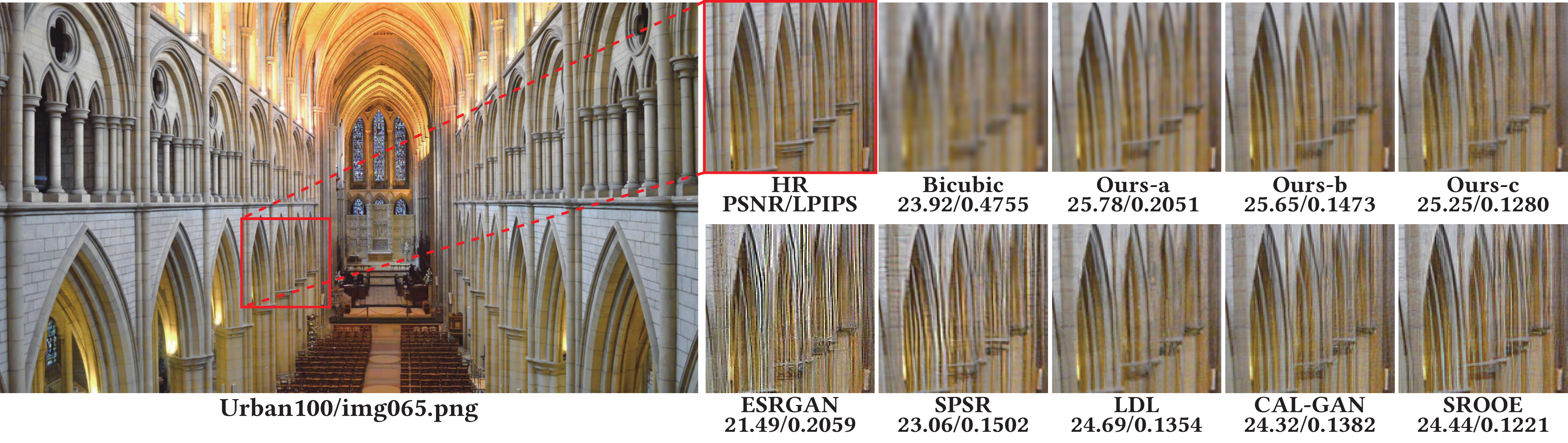}
		\caption{More visual comparisons on Urban100 \cite{Huang15}.}
		\label{fig:visual_compare_1}
	\end{figure*}
	
	\begin{figure*}[h]
		\centering
		\includegraphics[width=\textwidth]{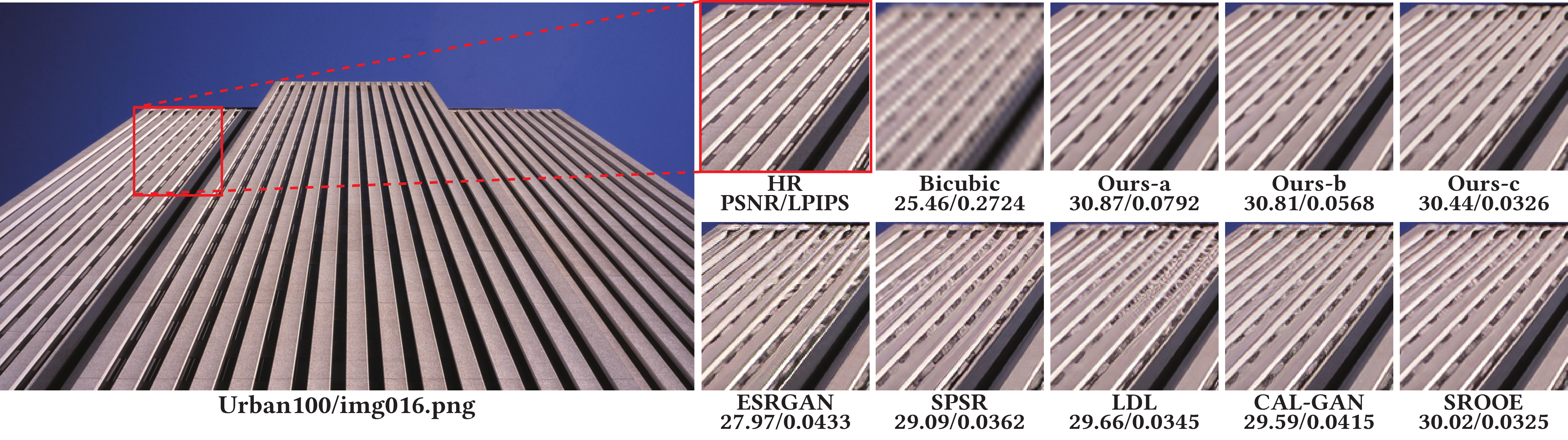}
		\includegraphics[width=\textwidth]{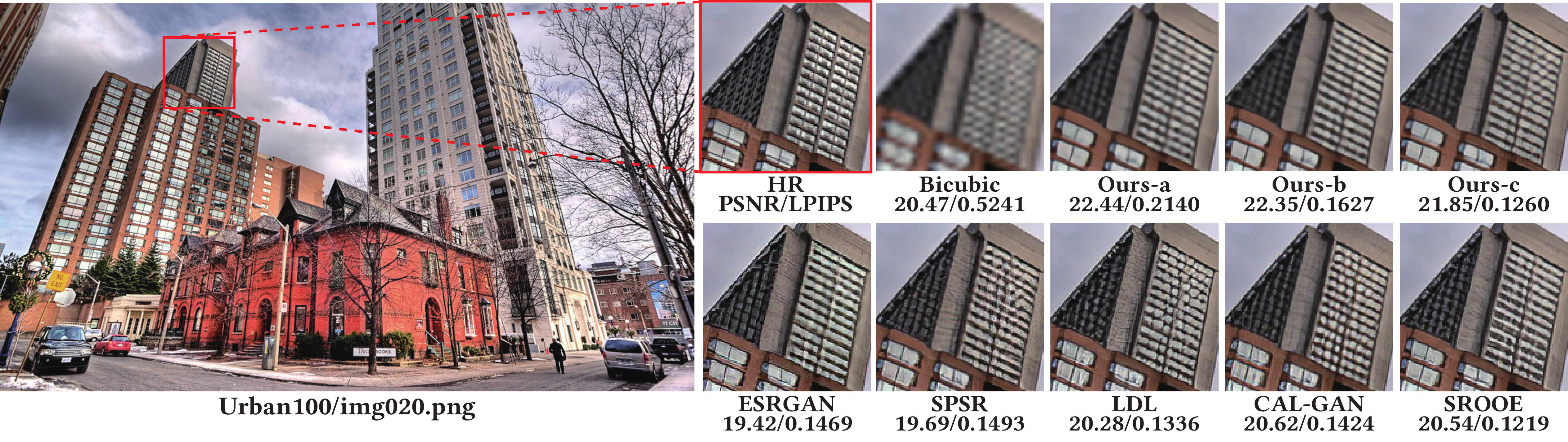}
		\includegraphics[width=\textwidth]{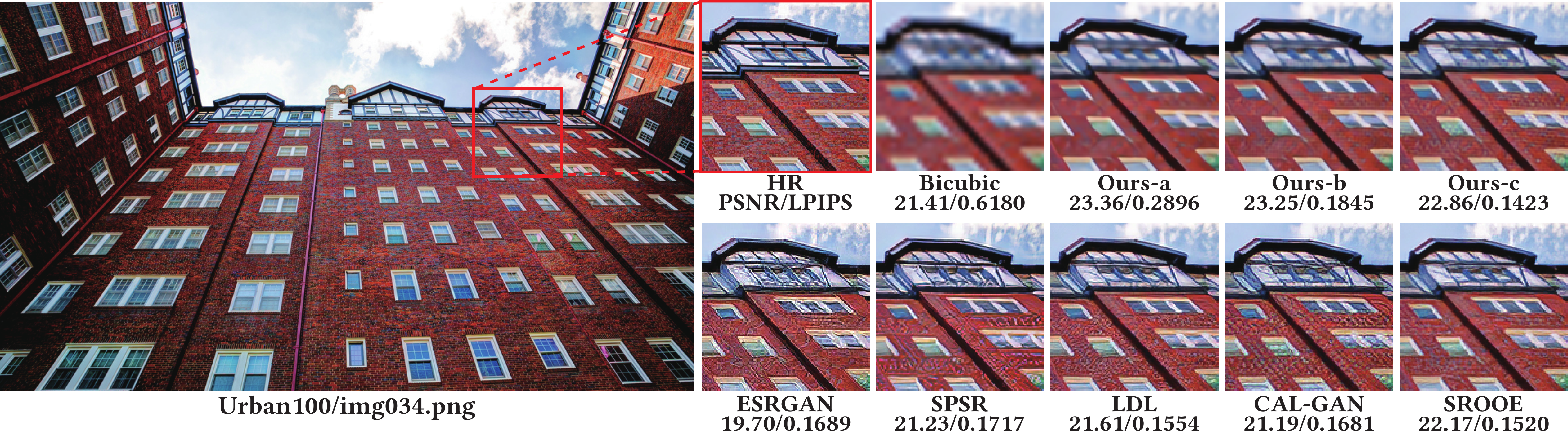}
		\includegraphics[width=\textwidth]{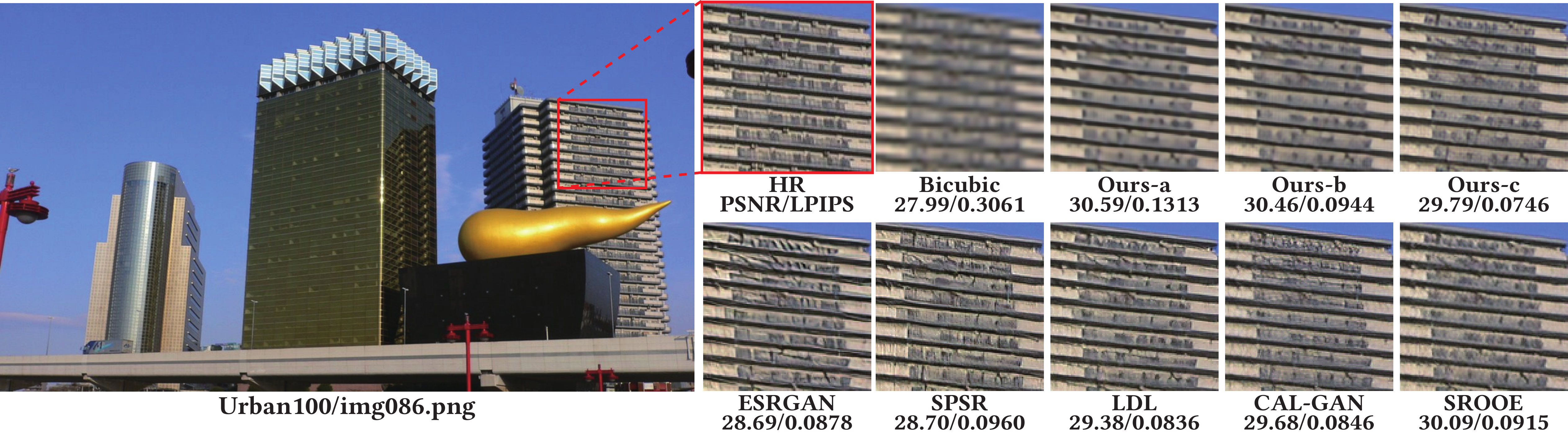}
		\caption{More visual comparisons on Urban100 \cite{Huang15}.}
		\label{fig:visual_compare_2}
	\end{figure*}
	\begin{figure*}[h]
		\centering
		\includegraphics[width=\textwidth]{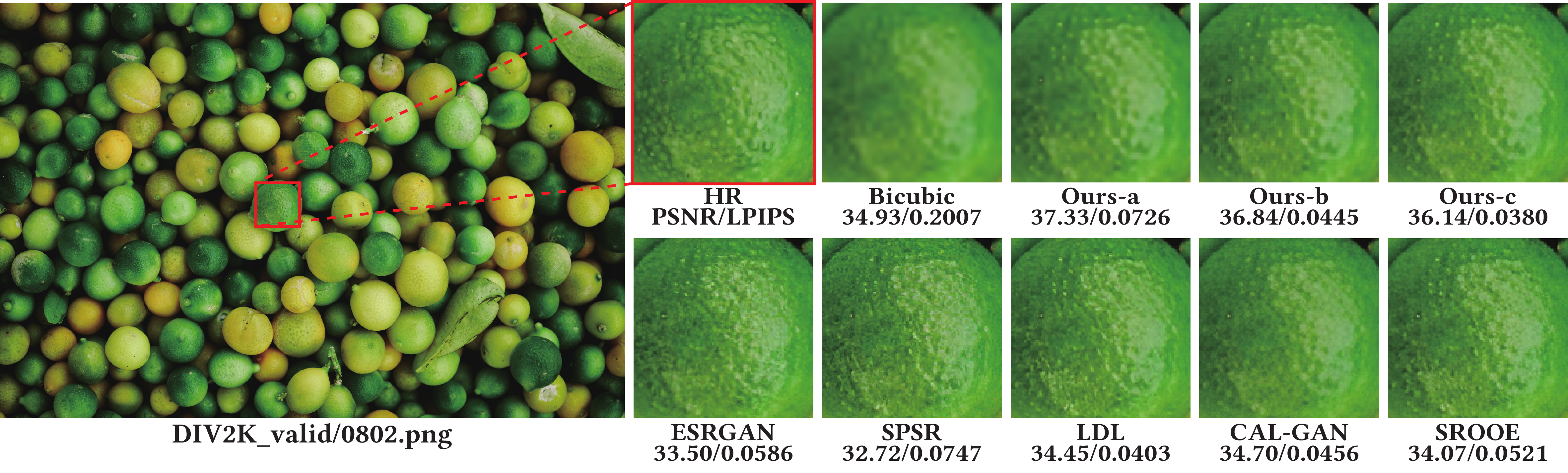}
		\includegraphics[width=\textwidth]{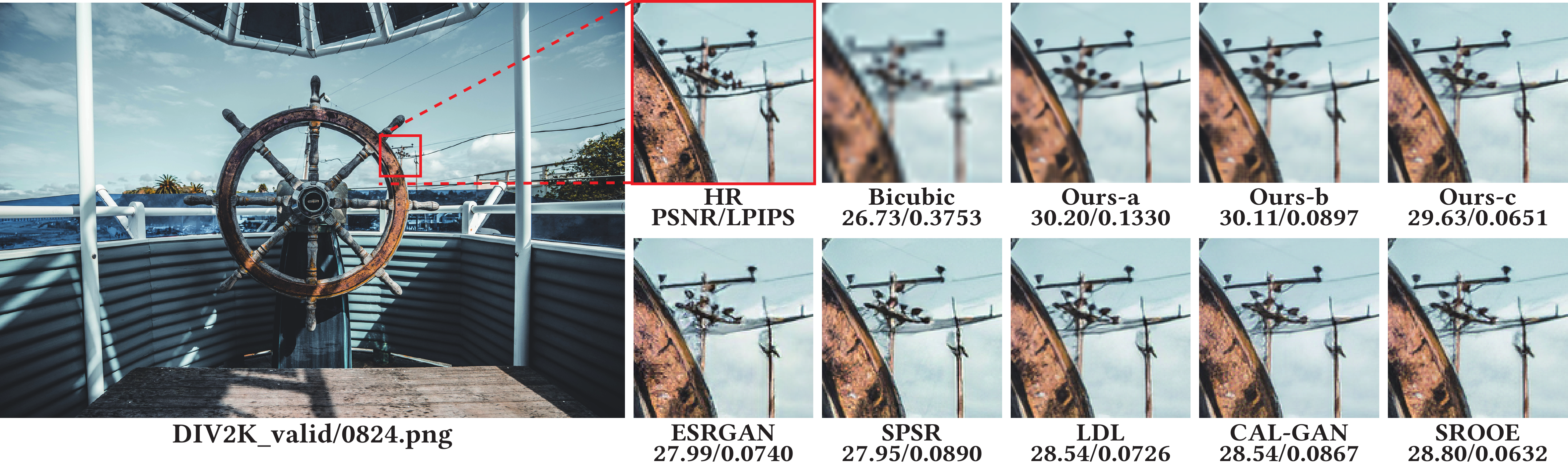}
		\includegraphics[width=\textwidth]{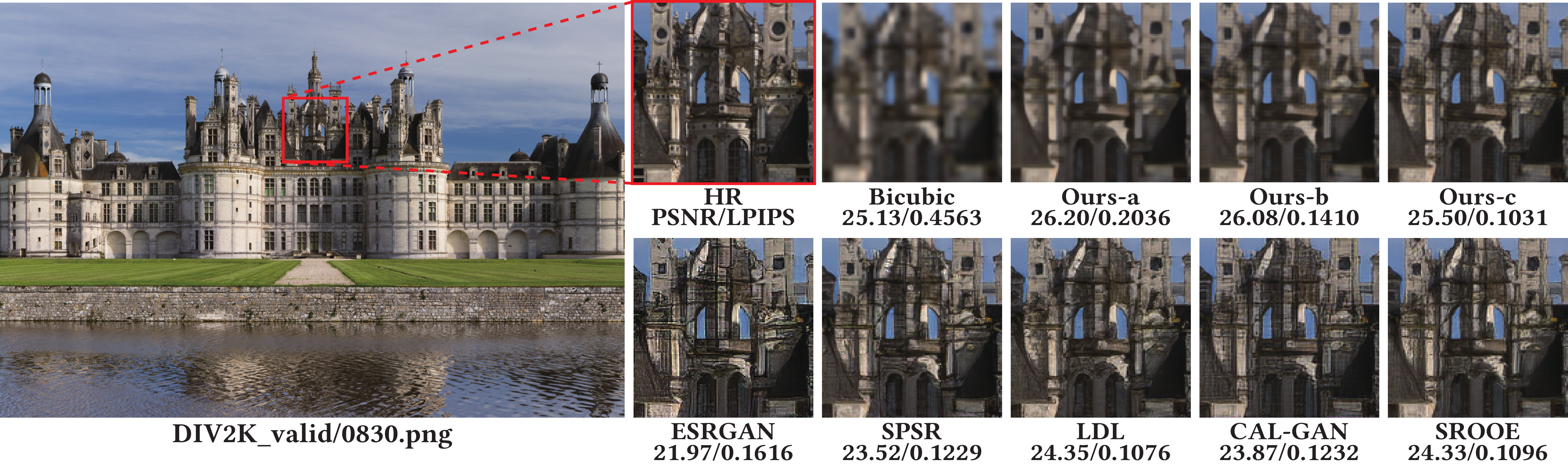}
		\includegraphics[width=\textwidth]{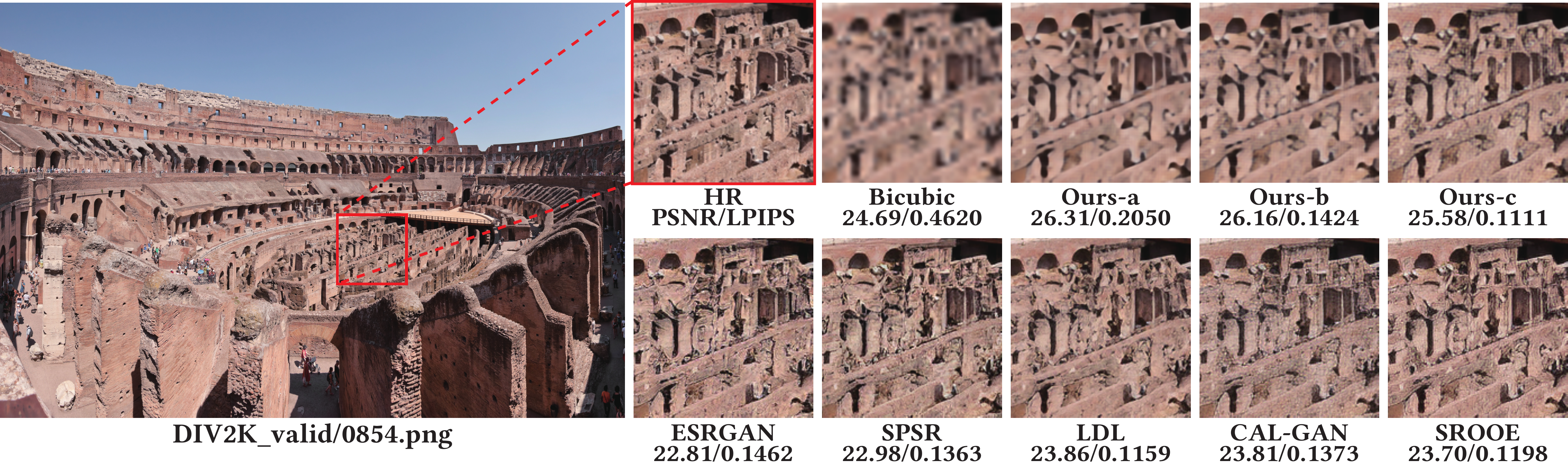}
		\caption{More visual comparisons on DIV2K \cite{Agustsson17} validation set.}
		\label{fig:visual_compare_3}
	\end{figure*}
	
	We have included more visual comparison results here, including the visualizations on the Urban100 \cite{Huang15} dataset as shown in Figure \ref{fig:visual_compare_1} and Figure \ref{fig:visual_compare_2}, and the visualizations on the DIV2K \cite{Agustsson17} validation set as shown in Figure \ref{fig:visual_compare_3}.
	
	\bibliographystyle{ACM-Reference-Format}
	\bibliography{refs}